%% file: output.tex
\documentclass[12pt]{article}
\usepackage{amsmath}
\usepackage{graphicx}
\usepackage{enumerate}
\usepackage{natbib}
\usepackage{url} 

\long\def\ignore#1{} 
\usepackage[inline,shortlabels]{enumitem}
\usepackage{float}
\usepackage{mathtools}
\usepackage{subdepth}
\usepackage{tikz}
\usetikzlibrary{arrows.meta}
\usetikzlibrary{decorations.markings}
\usepackage[linesnumbered]{algorithm2e}
\RestyleAlgo{ruled}

\tikzset{+ /.tip = {Bar[sep=-3pt 2,width=3pt 4]_[sep=0]}}
\usepackage[english]{babel}
\usepackage{longtable}
\usepackage{color}
\usepackage{mathtools}
\usepackage{amssymb,amsmath,amsthm}
\usepackage{multirow}
\usepackage[titletoc,title]{appendix}
\usepackage{authblk}
\usepackage{bbm}
\usepackage{setspace}
\usepackage{dsfont}
\usepackage[OT1]{fontenc}
\usepackage{subcaption}
\usepackage{refcount}
\usepackage{booktabs}

\usepackage{xr-hyper}
\usepackage[colorlinks,citecolor=blue,urlcolor=blue]{hyperref}

\newtheorem{remark}{Remark}

\newtheorem{theorem}{Theorem}
\AtEndDocument{\refstepcounter{theorem}\label{finalthm}}
\AtEndDocument{\refstepcounter{equation}\label{finaleq}}

{
  \theoremstyle{definition}
  \newtheorem{assumption}{}
}
{
  \theoremstyle{definition}
  
}
{
  \theoremstyle{definition}
  
}

\newcommand{\cate}{\mathsf{CATE}}
\newcommand{\ecate}{\mathsf{cCATE}}
\newcommand{\ate}{\mathsf{ATE}}
\newcommand{\eate}{\mathsf{cATE}}

\AtEndDocument{\refstepcounter{lemma}\label{finallemma}}

\newtheorem{definition}{Definition}

\renewcommand{\P}{\mathsf{P}}
\newcommand{\p}{\mathsf{p}}

\newcommand{\IF}{\mathsf{IF}}

\newcommand{\one}{\mathds{1}}
\newcommand{\E}{\mathsf{E}}
\renewcommand{\P}{\mathsf{P}}

\usepackage{accents}

\renewenvironment{proof}{{\it Proof }}{\qed \\}

\DeclarePairedDelimiterX{\norm}[1]{\lVert}{\rVert}{#1}

\pgfdeclarelayer{background}
\pgfsetlayers{background,main}
\usetikzlibrary{arrows,positioning}
\tikzset{
>=stealth',
punkt/.style={
rectangle,
rounded corners,
draw=black, very thick,
text width=6.5em,
minimum height=2em,
text centered},
pil/.style={
->,
thick,
shorten <=2pt,
shorten >=2pt,}
}
\newcommand{\Vertex}[3]
{\node[minimum width=0.6cm,inner sep=0.05cm] (#2) at (#1) {#3};
}
\newcommand{\Vertexr}[3]
{\node[rectangle, draw, minimum width=0.6cm,inner sep=0.05cm] (#2) at (#1) {#2};
}
\newcommand{\ArrowR}[3]%
{ \begin{pgfonlayer}{background}
\draw[->,#3] (#1) to[bend right=30] (#2);
\end{pgfonlayer}
}
\newcommand{\ArrowLW}[3]%
{ \begin{pgfonlayer}{background}
\draw[->,#3] (#1) to[bend left=30] (#2);
\end{pgfonlayer}
}

\newcommand{\ArrowL}[3]%
{ \begin{pgfonlayer}{background}
    \draw[->,#3] (#1) to[bend left=45] (#2);
  \end{pgfonlayer}
}

\newcommand{\EdgeL}[3]%
{ \begin{pgfonlayer}{background}
\draw[dashed,#3] (#1) to[bend right=-45] (#2);
\end{pgfonlayer}
}

\newcommand{\Arrow}[3]%
{ \begin{pgfonlayer}{background}
\draw[->,#3] (#1) -- +(#2);
\end{pgfonlayer}
}

\allowdisplaybreaks
\pgfarrowsdeclare{arcs}{arcs}{...}
{
  \pgfsetdash{}{0pt} 
  \pgfsetroundjoin   
  \pgfsetroundcap    
  \pgfpathmoveto{\pgfpoint{-5pt}{5pt}}
  \pgfpatharc{180}{270}{5pt}
  \pgfpatharc{90}{180}{5pt}
  \pgfusepathqstroke
}
\newcommand{\ArrowB}[3]%
{ \begin{pgfonlayer}{background}
    \draw[|-arcs,line width=0.4mm,shorten <= 0.3cm,shorten >= 0.3cm,#3] (#1) -- +(#2);
  \end{pgfonlayer}
}

\newcommand{\blind}{1}

\begin{document}

\def\spacingset#1{\renewcommand{\baselinestretch}%
{#1}\small\normalsize} \spacingset{1}


\if1\blind
{
  \title{\bf Identification and estimation of causal effects using non-concurrent controls in platform trials}
  \author{Michele Santacatterina\thanks{Corresponding author: santam13@nyu.edu. This article is based upon work supported by the National Science Foundation under
Grant No 2306556},
    Federico Macchiavelli Gir\'on, 
    Xinyi Zhang, and 
    Iv\'an D\'iaz \\
    Division of Biostatistics, Department of Population Health, \\ New York University School of Medicine, \\ 
    New York, NY, 10016}
  \maketitle
} \fi

\if0\blind
{
  \bigskip
  \bigskip
  \bigskip
  \begin{center}
    {\LARGE\bf Identification and estimation of causal effects using non-concurrent controls in platform trials}
\end{center}
  \medskip
} \fi

\bigskip
\begin{abstract}
Platform trials are multi-arm designs that simultaneously evaluate multiple treatments for a single disease within the same overall trial structure. Unlike traditional randomized controlled trials, they allow treatment arms to enter and exit the trial at distinct times while maintaining a control arm throughout. This control arm comprises both concurrent controls, where participants are randomized concurrently to either the treatment or control arm, and non-concurrent controls, who enter the trial when the treatment arm under study is unavailable. While flexible, platform trials introduce the challenge of using non-concurrent controls, raising questions about estimating treatment effects. Specifically, which estimands should be targeted
? Under what assumptions can these estimands be identified and estimated? Are there any efficiency gains? In this paper, we discuss issues related to the  identification and estimation assumptions of common choices of estimand. We conclude that the most robust strategy to increase efficiency without imposing unwarranted assumptions is to target the concurrent average treatment effect (cATE), the ATE among only concurrent units, using a covariate-adjusted doubly robust estimator. Our studies suggests that, for the purpose of obtaining efficiency gains, collecting important prognostic variables is more important than relying on non-concurrent controls. We also discuss the perils of targeting ATE due to an untestable extrapolation assumption that will often be invalid. We provide simulations illustrating our points and an application to the ACTT platform trial, resulting in a 20\% improvement in precision compared to the naive estimator that ignores non-concurrent controls and prognostic variables.
\end{abstract}

\noindent%
\vfill

\newpage
\spacingset{1.9} 

\section{Introduction}

\begin{figure}
    \centering
	\includegraphics[scale=0.50]{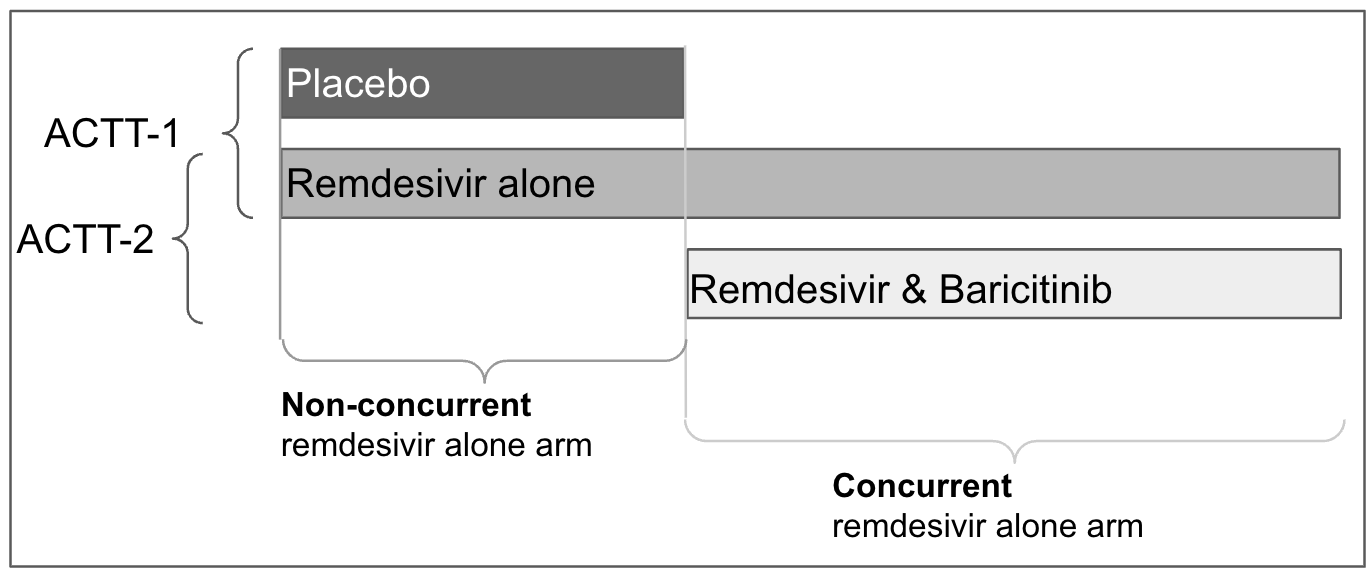}
	\caption{Adaptive COVID-19 Treatment Trial (ACTT) schema. Example of concurrent and non-concurrent arm}
	\label{fig:nonconc}
\end{figure}

Platform trials are multi-arm designs that simultaneously evaluate multiple treatments for a single disease within the same overall trial structure \citep{woodcock2017master,berry2015platform,park2022use}. Unlike traditional randomized controlled trials,  they allow treatment arms to enter and exit the trial at distinct times while maintaining a control arm throughout. These trials have been instrumental in assessing the efficacy of treatments across various therapeutic areas \citep[among others]{barker2009spy,foltynie2023towards,wells2012metacognitive} and gained traction during the COVID-19 pandemic \citep[among others]{hayward2021platform,angus2020effect,kalil2021baricitinib}. For instance, the Adaptive COVID-19 Treatment Trial (ACTT) \citep{kalil2021baricitinib} was a platform trial that investigated treatments for hospitalized adult patients with COVID-19 pneumonia. ACTT comprised of multiple stages, as depicted in Figure \ref{fig:nonconc}. In the initial stage (ACTT-1), the efficacy of remdesivir alone versus placebo was evaluated. Subsequently, in the second stage (ACTT-2), placebo was discontinued, and a new treatment, remdesivir plus baricitinib, was introduced while concurrently randomizing participants to remdesivir alone. Here, the remdesivir alone arm served as a shared
arm between the ACTT-1 and ACTT-2 stages. The remdesivir alone arm is termed \textit{non-concurrent} for remdesivir plus baricitinib during ACTT-1 and \textit{concurrent} during ACTT-2. In this paper, we adhere to the terminology used in current literature \citep{bofill2023use,lee2020including} and designate the shared arm as \textit{control}, irrespective of whether it is a placebo arm or an active control or an experimental treatment. Thus, we consistently use the terms concurrent and non-concurrent controls regardless of the nature of the shared arm.

The central question revolves around the efficient utilization of non-concurrent controls to estimate treatment effects in platform trials. Specifically, \textit{what estimands should be targeted to evaluate the causal effect of a treatment versus a shared control? Under what assumptions can these estimands be identified and estimated? Does using non-concurrent controls lead to efficiency gains?} 

Addressing these questions requires careful consideration of how the timing of entry into the platform trial may introduce bias into the study results, which is referred to as ``time drift'', ``temporal drift'' or ``time trend''. Various methods have been proposed to control for it, including test-then-pool approaches \citep{Viele2014}, frequentist and Bayesian regression models \citep{lee2020including,sridhara2022use,bofill2023use,saville2022bayesian}, propensity-score-based methods \citep{Yuan2019,Chen2020}, and other approaches \citep{Han2017,Collignon2020,Ibrahim2000,Neuenschwander2009,Banbeta2019,gravestock2017adaptive,Bennett2021,Hobbs2011,Normington2020,Schmidli2020,Hupf2021,jiang2023elastic}.

While these methods provide a statistical way to incorporate non-concurrent controls and control for the ``temporal drift'' bias, these approaches are ``model-first'', meaning that they are focused on first providing a model for the outcome and then reverse engineering interpretations for the estimated parameters in terms of causal effects. This approach conflicts with the recently advocated \textit{estimand framework} \citep{food2021e9} and the International Council for Harmonisation (ICH) E9(R1) guidance \citep{international2017harmonised}, where the causal target of interest is first identified based solely on scientific discussions, and then the optimal statistical estimation method for that target is deployed. Existing methods for the use of non-concurrent controls lack a formal framework for characterizing causal effects and their identifying conditions, which implies that interpreting the effect estimates from these procedures and making recommendations regarding clinical practice become challenging. These concerns are underscored in recent reviews \citep{collignon2022estimands,koenig2024current} and are discussed in the FDA estimand framework \citep{food2021e9} and the ICH E9(R1) guidance \citep{international2017harmonised}.

In this paper, we discuss the use of non-concurrent controls, using the estimand framework to guide our discussion and choices. We 
propose to target the concurrent average treatment effect of treatment arm $k$, $\eate(k)$, as an estimand of interest in platform trials. Specifically, $\eate(k)$ is the marginal average difference in outcomes for individuals who receive treatment $k$ compared to those in the shared control group, among the concurrent population. We then provide assumptions for its non-parametric identification, and show that these assumptions are all feasible, in contrast to the assumptions required for identification of the average treatment effect, which are not testable. We develop several estimators for $\eate(k)$, including outcome regression, inverse probability weighting, and doubly robust estimators. We show that efficiency gains can be obtained by leveraging non-concurrent controls for estimators based on outcome regression under correct models specification. Interestingly, we also show that there are no asymptotic gains in efficiency when using non-concurrent controls with doubly robust estimators adjusted by time of entry into the trial when treatment availability is a deterministic function of entry time. However, we show that efficiency gains can be obtained when treatment availability is a stochastic function of entry time.

In randomized trials, efficiency gains may come from multiple sources. For instance, one can attempt to gain efficiency by increasing the sample size, as illustrated by the use of non-concurrent controls. Alternatively, precision may be increased through adjustment for prognostic variables \cite[see][among others]{colantuoni2015leveraging,benkeser2021improving}. 
Prognostic variables are often incorporated through regression models, which can then be mapped into conditional or marginal effect estimates. However, it is important to remember that outcome models may lead to biased results under certain types of misspecification. It is therefore important to use doubly robust estimators for covariate adjustment.  Doubly robust estimators are consistent when either the treatment assignment or the outcome model is correctly specified, a property we obtain ``for free'' in platform trials due to randomization. 
Therefore, a key takeaway when targeting $\eate(k)$ in platform trials is to use a doubly robust estimator that prioritizes identifying strong prognostic baseline variables rather than relying on non-concurrent controls. The latter provides no efficiency benefit when using a robust estimator that does not rely on the ability to correctly specify the outcome regression mechanism when treatment availability is a deterministic function of entry time.

Finally, we further highlight the risks of targeting the average treatment effect \textcolor{black}{ of treatment $k$ ($\ate(k)$)} using data for the entire duration of the trial including the non-concurrent period, due to its dependence on an untestable extrapolation assumption. \textcolor{black}{The $\ate(k)$ is the  marginal average difference in outcomes for individuals who receive treatment $k$ compared
to those in the shared control group in the entire trial population (non-concurrent and concurrent units).}

\section{Notation and setup}

For each of $i\in\{1,\ldots,n\}$ study participants, let $E_i$ denote the (random) entry time, after eligibility screening and consent, of a unit into the study, let $W_i$ denote a set of baseline variables, let $A_i$ denote the randomized treatment taking values $k= 0,1,\ldots,K$, where $0$ denotes the control arm and $1,\ldots,K$ denotes the treatments of interest. Let $V_{k,i}$ denote an indicator of whether arm $k$ was available at time $E_i$, and define $V_i=(V_{0,i},\ldots, V_{K,i})$. Let $Y_i$ denote a binary or numerical outcome measured at a fixed time after entry $E_i$. The observed data is $D=(Z_1,\ldots,Z_n)$, where $Z_i$ represents the data for the experimental unit $i$, \textit{i.e.},
$Z_i=(E_i, W_i, V_i, A_i, Y_i)\sim\P$. 
\textcolor{black}{We define $V_{0,i}=\cdots=V_{k,i}=1$ with probability one so that at least $k\ge 1$
treatments plus control are available at the start of the trial.}
We also assume the data are ordered based on the time of study entry in the sense that $E_1\leq E_2\leq \cdots \leq E_n$. 
\textcolor{black}{Note that $\P(A_i=k\mid V_{k,i}=0)=0$ by design.}

\subsection{A structural causal model and associated DAG}


To encapsulate the role of entry time and non-concurrent controls in platform trials, we posit the structural causal model and directed acyclic graph (DAG) \citep{pearl1995causal} represented in  Figure \ref{fig:dag-b}, and its interpretation in terms of a non-parametric structural equation model in eq. (\ref{eq:npsem}) respectively.

\begin{minipage}{0.45\textwidth}
    \begin{figure}[H]
        \small
        \centering
        \input{dag-b.tex}
        \caption{DAG associated to the structural equation model in equation    (\ref{eq:npsem}).}     \label{fig:dag-b}
    \end{figure}
\end{minipage}%
\hspace{6mm}
\begin{minipage}{0.45\textwidth}
    \begin{figure}[H]
        \small
        \centering
        \begin{align}
          E_i&=f_E(U_{E,i}),\notag\\
          W_i&=f_W(E_i, U_{W,i}),\notag\\
          V_{k,i}&= f_{V_k}(E_i,U_{V_k,i}),\label{eq:npsem}\\
          A_i&=f_A(V_{k,i}, W_i, U_{A,i}),\notag\\
          Y_i&=f_Y(A_i, W_i, E_i, U_{Y,i}).\notag
        \end{align}
    \end{figure}
\end{minipage}%

\vspace{0.2in}

\noindent 
We now discuss some important features of Model~(\ref{eq:npsem}).  Model~(\ref{eq:npsem}) allows all variables to be dependent, directly or through other variables, on entry time $E_i$, and therefore appropriately models temporal drifts.
It also allows the treatment assignment $A_i$ to depend on the participant's covariates $W_i$, thus allowing study designs such as stratified randomization \citep{broglio2018randomization}. Model~(\ref{eq:npsem}) also imposes some exclusion restrictions. 
First, the treatment assignment $A_i$ is not allowed to depend on the entry time $E_i$ other than through treatment availability $V_{k,i}$. In other words, a participant entering the study at time $E_i$ can only be assigned to available treatments at that time, but the randomization probability of a treatment that is available for assignment does not vary in time.  Second, the outcome for unit $i$, $Y_i$, is not allowed to depend on the availability of treatments $V_{k,i}$, other than through the treatment actually given to unit $i$, but it is allowed to directly depends on unit's $i$ entry time.  Third, the availability of treatments $V_{k,i}$ does not depend on covariates $W_i$. 
The two last assumptions are reasonable since the treatments under evaluation do not often depend on trial data. 
In this paper, we assumed that covariates $W_i$ depend on entry time $E_i$, and not the other way around. This assumption is reasonable because, in many platform trials, entry time does not depend on individual-level data. Furthermore, it can be shown that the results presented in the next sections also hold when entry time $E_i$ depends on $W_i$.
\textcolor{black}{In addition, in this paper, we assumed that the availability of treatment $k$, $V_{k,i}$, depends on the entry time $E_i$, and not the other way around. However, it is possible that entry time could depend on the availability of certain treatment arms, for example, if some treatments are more attractive. We demonstrate in the Appendix that the identification of  $\eate(k)$ under these conditions remains consistent with that under Model~(\ref{eq:npsem}).}
In Model~(\ref{eq:npsem}), the functions $f_E$, $f_W$, and $f_Y$ are completely unknown, thus making the model non-parametric, while the treatment assignment function, $f_A$, and treatment availability function, $f_{V_{k}}$, are known by design.  In addition, the random variables $U_{E,i}$, $U_{W,i}$, and $U_{Y,i}$ are unmeasured factors that impact the entry time, covariates, and outcomes, respectively. The random variables $U_{A,i}$ control the randomization probabilities and are known by design. The random variables $U_{V_{k}}$ represent all factors that determine the availability of treatments for subjects in the trial. 

In the following section, under this model and its associated DAG, we define the concurrent average treatment effect as the causal estimand of interest and introduce its identification assumptions, aligning with the estimand framework advocated by the FDA \citep{food2021e9}.

\section{Definition and identification of the concurrent average treatment effect}\label{sec:identification}

In this paper, we focus on endpoints measured at fixed time-points post-randomization. 
Additionally, we consider an intention-to-treat (ITT) analysis. Our results can be easily extended to binary endpoints. We define the concurrent average treatment effect of treatment $k$ against a shared control arm in terms of counterfactual variables \citep{Pearl10},
$Y_i(k)=f_Y(k, V_{k,i}, W_i, E_i, U_{Y,i})$, where $k=0,\dots, K$, that would have been observed
in a hypothetical world where treatment $A_i=k$ had been given, \textit{i.e.},
$\P(A_i=k)=1$. We first define it and then
discuss their non-parametric identification.

\begin{definition}[Conditional and marginal average treatment effect of treatment
  arm $k$ compared to shared control among concurrent population]
  \begin{align*}
    \ecate(k,w,e) &= \E[Y(k) - Y(0)\mid W=w, E=e,V_k=1]\\
    \eate(k) &= \E[\ecate(k,W,E)\mid V_k=1].
  \end{align*}
\end{definition}

\noindent
 $\eate(k)$ is the ITT-average treatment effect among \textit{only} concurrent units, $V_k=1$. $\ecate(k,w,e)$ is its conditional versions, conditioning on baseline variables $W$ and entry time $E$. We now provide assumptions to identify it.


\subsection{Non-parametric identification}
\label{identification}
Non-parametric identification allows us to express the causal target
quantity of interest in terms of the distribution of the observed data
without relying on assumptions on the functional form of the
distributions \citep{pearl1995causal}. 
In order to discuss non-parametric identification of $\eate(k)$, we introduce the following assumptions: 

\color{black}

\begin{assumption}[weak A-ignorability]\label{ass:wA-ignorability}  
    $\E[Y(k) | W=w,E=e,V_k=1] = \E[Y(k) | A=k,W=w,E=e,V_k=1]$.
\end{assumption}

\noindent
Assumption~\ref{ass:wA-ignorability} is an untestable assumption, \textit{i.e.,} it is a function of counterfactuals which are unobservable; that state that once we control for $W$, $E$ and $V_k$, the counterfactual outcome under $k$ is independent from the treatment assignment. We expect this to hold by design because of randomization. 


\begin{assumption}[Consistency]\label{ass:cons} 
$\P(Y(k) | A=k, W=w, E=e, V_k=1) = \P(Y | A=k, W=w, E=e, V_k=1)$.
\end{assumption}

\noindent
Assumption~\ref{ass:cons} is a standard causal inference assumption that states that under $V_k=v$ and once controlled for $W, E$, the distribution of the observed outcome under $A=k$ is the same as that of the counterfactual outcome $Y(k)$ for all $k$ in $\{0,\ldots,K\}$. This assumption is implied by the structural causal model (\ref{eq:npsem}). We expect this also to hold by design. 

\begin{assumption}[Positivity of treatment assignment mechanism among
  concurrent units]\label{ass:posa}
$\P(A=k\mid W=w, E=e, V_k=1)>0$ for all $w$ and $e$ s.t. $V_k=1$.
\end{assumption}

\noindent
Assumption~\ref{ass:posa} states that once a treatment arm is available in the trial all covariate profiles $w,e$ have a positive probability of receiving such treatment.

\begin{assumption}[Positivity of shared arm assignment mechanism among
  all controls]\label{ass:posa0}
  $\P(A=0\mid W=w, E=e)>0$ for all $w$ and $e$.
\end{assumption}

\noindent
Similarly to Assumption~\ref{ass:posa}, Assumption~\ref{ass:posa0} states that within the shared control arm, covariate profiles $w,e$ have a positive probability of receiving the control group (Note that assumption~\ref{ass:posa0} is redundant given assumption~\ref{ass:posa} but helps clarifying our identification proofs). These two assumptions hold by design in platform trials. 

\begin{assumption}[Pooling concurrent and non-concurrent controls]\label{ass:condex}
   Assume \\ $\E(Y\mid A=0, W=w, E=e, V_k=1) = 
  \E(Y\mid A=0, W=w, E=e)$ for all $e$ s.t. $V_k=1$.
\end{assumption} 

\noindent
Assumption~\ref{ass:condex} states that once we control for $W$ and $E$, the conditional expectation of outcome $Y$ under control ($A=0$) in the pooled concurrent and non-concurrent units ($V_k=0$, and $V_k=1$ -- right-hand side of assumption~\ref{ass:condex}) is the same as that among only concurrent units ($V_k=1$ -- left-hand side of assumption~\ref{ass:condex}), for all values $e$ among the concurrent units. In other words, after conditioning on $W$ and $E$, what is learned using all the pooled data can be used to predict conditional expectations under only concurrent. In addition, it is straightforward to see that these quantities depend only on \textit{observable} data. For instance, we know by design that we have data for the left-hand side of assumption~\ref{ass:condex} for all $e$ such that $V_k=1$ and for the right-hand side, for all $e$ in the shared arm. Therefore assumption~\ref{ass:condex} can be tested as discussed in our practical guidelines in section \ref{sec:pract_guide}.  

\color{black}


\noindent

\begin{remark}
    Under Model (\ref{eq:npsem}) with $V_{k}$ a deterministic function of $E$, \textit{i.e.}, $V_{k,i} = \one[{E_i > t_k}]$, where $t_k$ is a positive scalar, assumption~\ref{ass:condex} holds always true at the population level and for the true conditional outcome expectations. \label{rem:condex}
\end{remark}

\noindent
In contrast, when $V_{k,i}$ is a stochastic function of $E$, \textit{i.e.,} $V_{k,i}=f_{V_k}(E_i,U_{V_k,i})$, an assumption is needed since we do not know $U_{V_k,i}$. \textit{e.g.,} the unknown error can be different between the two expectations. 
In addition, while this remark is always true at the population level and for the true conditional outcome expectations, for given estimators, its validity depends on the correct model specification. For instance, if there are non-linearities in $W$ in the data, and we fit linear models within the pooled dataset and within the concurrent dataset, these two linear regressions will not be equal because they will capture the projection of the true non-linear expectation onto linear models in different subset of the range of $W$. This underscores the importance of using non-parametric models for these regressions if data are to be pooled.  




We now provide an identification theorem, under assumptions \ref{ass:wA-ignorability}-\ref{ass:condex}.

\begin{theorem}[Identification of $\eate$ in
  platform adaptive trials under Model~(\ref{eq:npsem})]\label{theo:modela}
  Assume Model~(\ref{eq:npsem}) and assumptions \ref{ass:wA-ignorability}-\ref{ass:posa}. Then we have:
  \begin{enumerate}
      \item The parameter
  $\ecate(k,w,e)$ is identified as \label{stat:1}
  \begin{equation}
    \label{eq:ecatemodela}
    \E(Y\mid A=k,W=w,E=e,V_k=1) - \E(Y\mid A=0, W=w,E=e, V_k=1),
\end{equation}
\item Under assumptions \ref{ass:wA-ignorability}-\ref{ass:condex}, $\ecate(k,w,e)$ is also
identified as \label{stat:2}
  \begin{equation}
    \label{eq:ecateexI}
    \E(Y\mid A=k, W=w,E=e, V_k=1) - \E(Y\mid A=0, W=w, E=e).    
  \end{equation}
  \end{enumerate}
  Furthermore, $\eate(k)$ is identified by taking the average of the
  above expression for $\ecate(k)$ over the distribution of $(W,E)$ conditional on
  $V_k=1$.
\end{theorem}


\noindent
Equivalent expressions based on weighting are provided in the appendix. A comparison between expressions (\ref{eq:ecatemodela}) and
(\ref{eq:ecateexI}) reveals why researchers have been historically motivated to use non-concurrent
controls: they can be useful in estimating the outcome expectation for
the controls, therefore potentially reducing the variance of the estimator.

In the next sections, we will show that, when $V_{k}$ is a deterministic function of $E$, efficiency gains only bear out for plug-in estimators based on parametric regressions, which can be biased if the models are misspecified.
Doubly robust estimators, which are always consistent by virtue of randomization, will not benefit from these efficiency gains. However, we will also show, that efficiency gains can be obtained when $V_{k}$ is a stochastic function of $E$. 

\subsection{On the identification of the average treatment effect}
\label{sec:ident_ATE}

In this paper, we propose targeting $\eate(k)$ in platform trials. However, many researchers are familiar with another estimand, the average treatment effect, defined as the expected difference \textcolor{black}{in outcomes} between treatment $k$ and control $0$ in the entire trial population ($V_k=1$ and $V_k=0$). In formulas,

\begin{definition}[Conditional and marginal average treatment effect of treatment
  arm $k$ compared to shared control]
  \begin{align*}
    \cate(k,w,e) &= \E[Y(k) - Y(0)\mid
    W=w,E=e]\\
    \ate(k) &= \E[\cate(k,W,E)],
  \end{align*}
\end{definition}

\noindent
where $\cate(k,w,e)$ is the conditional version of $\ate(k)$, conditioning on baseline variables $W$ and entry time $E$. The familiarity of $\ate(k)$ is partly because, in standard randomized controlled trials, the common statistical model used to evaluate treatment effects is a linear model that regresses the outcome on the treatment group and baseline covariates. In such model, the canonical interpretation of the model coefficient for the treatment aligns with that of the $\ate(k)$. For instance, assuming the following linear model: $\E \left[ Y | A, W, E \right] = g(A, W, E; \alpha) = \eta + \beta A + W \gamma_w + E \gamma_e$, we can show that,
\begin{align*}
\ate(1) = \E[Y(1)-Y(0)] &= \E[\E[Y | A=1, W, E]] - \E[\E[Y | A=0, W, E]] \\
&= \E[ \eta + 1 \beta  + W\gamma_w + E\gamma_e ] - \E[ \eta + 0 \beta  + W\gamma_w + E\gamma_e ] \\
&= \beta.
\end{align*}
\noindent
We now explain why we believe that targeting $\ate(k)$ in platform trials is dangerous. To do so, we start by introducing an additional assumption needed to identify $\ate(k)$ in platform trials: 

\begin{assumption}[Extrapolation of outcome mechanism
  among the treated]\label{ass:extra} Assume \\
  $\E(Y \mid A=k, W=w, E=e, V_k=1)= \E(Y\mid A=k, W=w, E=e)$ for all $e$.
\end{assumption}

\noindent
Assumption~\ref{ass:condex} and assumption \ref{ass:extra} state that the outcome distribution among controls and treated are exchangeable between patients for whom treatment $k$ is available and those for whom it is not, respectively; given patients' baseline variables and entry time. Note that assumption \ref{ass:condex} and assumption \ref{ass:extra} are similar in nature in that they assume exchangeability of the outcome mechanism for treatment and control arms. However, 
there is a fundamental difference between these assumptions that makes identification based on \ref{ass:condex} more reliable than identification based on \ref{ass:extra}. Assumption~\ref{ass:condex} is a testable assumption since it is based on observed data.  In addition, as aforementioned, assumption \ref{ass:condex} is a statement that holds always true when $V_k$ is a determinist function of $E$ and its validity only depends on the correct model specification. 
Consequently, whether pooling data in a specific regression algorithm is appropriate can be empirically checked as shown in our practical guidelines. 

On the other hand, assumption \ref{ass:extra} is an identification assumption based on unobserved data, since it requires assuming that the conditional outcome expectation observed in patients who could hypothetically be randomized to treatment $k$ can be used to \textit{extrapolate} to those who could not. In other words, \ref{ass:extra} is an extrapolation assumption, since it assumes that the expected outcome under treatment $A=k$ in times $E=e$ and baseline variables $W=w$ of no treatment availability $V_k=0$ can be extrapolated from a model fit on times $E=e$ and baseline variables $W=w$ where the treatment was available. Consequently, assumption \ref{ass:extra} cannot be empirically checked.

We know state and show in the appendix that identification of $\ate(k)$ depends on the extrapolation assumption \ref{ass:extra}.


\begin{theorem}[Identification of $\ate(k)$ in
  platform adaptive trials under Model~(\ref{eq:npsem})]\label{theo:modela2}
  Assume Model~(\ref{eq:npsem}) and assumptions \ref{ass:wA-ignorability}-\ref{ass:extra}. Then we have that
  $\cate(k,w,e)$ is identified as (\ref{eq:ecateexI}). 
  Consequently, $\ate(k)$ is identified by taking the average of the
  above expression for $\cate(k)$ over the marginal distribution of $(W,E)$.
\end{theorem}

\noindent
In summary, unlike $\eate(k)$, $\ate(k)$ depends on an extrapolation assumption (\ref{ass:extra}) which can be risky. Firstly, this assumption cannot be tested. Secondly, it is often unrealistic for novel diseases with a rapidly changing pathology and clinical landscape.

\section{Relation to analytical approaches common in the literature}
\label{sec:relation}

Regression models are often used to estimate $\E(Y \mid A = k, E = e, V_k = 1)$ or $\E(Y \mid A = k, W = w, E = e, V_k = 1)$ and then used to extrapolate to units where $V_k = 0$, thus targeting $\ate(k)$ as discussed in Section \ref{sec:ident_ATE}.
Inferences are then made using the regression coefficient related to the treatment, whether within the frequentist \citep{lee2020including,BofillRoig2022a} or Bayesian framework \citep{saville2022bayesian,BofillRoig2022b,Ibrahim2000}. 

Matching techniques have been proposed to estimate the average treatment effect among the treated, $\E[ Y(k) - Y(0) | A=k ]$. The idea is to balance covariates $W$ between concurrent and non-concurrent controls by using for instance a matching algorithm based on the propensity score \citep{Yuan2019}. 

\ignore{
It is worth noticing that regression methods seem to assume a structural causal model described by the DAG in Figure \ref{fig:dag-reg}, while matching techniques follow that represented by the DAG in Figure \ref{fig:dag-weig}. In the first, entry time is considered as a common cause of both $A$, the treatment and $Y$ the outcome. In the second, the role of entry time $E$ is summarized by the selection into the concurrent population $V$ which is assumed to depend only on $W$. This is different from our Model~\ref{eq:npsem} where we assumed that the availability of treatment $k$ is not a function of individual level characteristics. 

\begin{minipage}{0.45\textwidth}
    \begin{figure}[H]
        \small
        \centering
        \input{dag-regression}
        \caption{DAG assumed in regression methods.}     \label{fig:dag-reg}
    \end{figure}
\end{minipage}%
\hspace{6mm}
\begin{minipage}{0.45\textwidth}
    \begin{figure}[H]
        \small
        \centering
        \input{dag-weight}
        \caption{DAG assumed in weighting methods.}     \label{fig:dag-weig}
    \end{figure}
\end{minipage}%
}

Bayesian methods have been proposed to include non-concurrent controls. The idea is to learn a prior of the parameter of interest using non-concurrent controls only. Then, this prior is combined with the concurrent control data via Bayes' theorem. Meta-analytic priors \citep{schmidli2014robust} or elastic priors \citep{jiang2023elastic} have been proposed.  These methods assume an exchangeability assumption for the control parameters, which relates to  \ref{ass:condex}. Other Bayesian methods have been proposed \citep[among others]{Neuenschwander2009,Bennett2021,wei2024bayesian}. These methods, however, do not allow for the use of baseline covariates $W$ and it is not clear what estimands they target.

In this paper, to estimate $\eate(k)$, we propose estimators based on outcome regression (OR) and inverse-probability-weighting (IPW), and doubly robust estimators. 

\section{Estimation of concurrent average treatment effect}
\color{black}

To build intuition, we start by introducing outcome regression (OR) and inverse probability weighting (IPW) estimators for $\eate(k)$ considering $V_k$ a deterministic function of $E$. Since OR and IPW estimators are not robust to model misspecification, we then propose doubly robust (DR) estimators.  To simplify notation, the following sections assume there are only two treatment arms $k=1$ and $k=0$. Furthermore, we assume that $V_1=\one\{E>t\}$ for some time $t$ such that treatment $A=1$ is only available for patients who entered the trial after time $t$.

\subsection{Estimators based on parametric outcome regression}
\label{sec:est_OR}
Based on the identification results presented in Theorem \ref{theo:modela}, eq.~(\ref{eq:ecatemodela}),  
the conditional mean $\E(Y\mid A=k, W=w,E=e,V_1=1)$, where $a=\lbrace 0,1 \rbrace$, can be modelled as 
$$\E(Y\mid A=k,W=w,E=e,V_1=1) = \mu_{\text{oc}}(k,w,e,1;\beta_a),$$
\noindent
where (oc) stands for only-concurrent. Based on the identification results presented in Theorem \ref{theo:modela}, eq.~(\ref{eq:ecateexI}) 
, 
the conditional mean $\E(Y\mid A=0, W=w,E=e)$, can be modelled as 
$$\E(Y\mid A=0, W=w,E=e) = \mu_{\text{all}} (0,w,e;\alpha_a).$$
\noindent
An estimate of $\beta_a$ and $\alpha_a$ can be then obtained by using maximum likelihood estimation, \textit{i.e.}, ordinary least squares, only among the concurrent controls $V_k=1$ for $\mu_{\text{oc}}(a,w,e,1;\beta_a)$ and among all concurrent and non-concurrent controls when using $\mu_{\text{all}}(0,w,e;\alpha_a)$, \textit{i.e.}, among $A=0$ only. Let $\hat{\beta}_a$ and $\hat{\alpha}_a$ denote consistent estimators of $\beta_a$ and $\alpha_a$, respectively. We then propose
\begin{align*}
    \hat{\eate}_{\text{OR}}^{\text{oc}} &= \frac{\sum_{i=1}^n \one{\{V_{k,i}=1}\} \mu_{\text{oc}}(1,w_i,e_i,1;\hat\beta_1)}{ \sum_{i=1}^n \one{\{v_{k,i}=1}\}  } \\ & - \frac{\sum_{i=1}^n \one{\{V_{k,i}=1}\} \mu_{\text{oc}}(0,w_i,e_i,1;\hat\beta_0)}{ \sum_{i=1}^n \one{\{V_{k,i}=1}\}  } ,
\end{align*}
\noindent
as an outcome regression estimator for $\eate(k)$. Under Theorem \ref{theo:modela}, eq.~(\ref{eq:ecateexI}), we propose the alternative outcome regression estimator for $\eate(k)$,
\begin{align*}
    \hat{\eate}_{\text{OR}}^{\text{all}} &= \frac{\sum_{i=1}^n \one{\{V_{k,i}=1}\} \mu_{\text{oc}}(1,w_i,e_i,1;\hat\beta_1)}{ \sum_{i=1}^n \one{\{v_{k,i}=1}\}  } \\ &- \frac{\sum_{i=1}^n \one{\{V_{k,i}=1}\} \mu_{\text{all}}(0,w_i,e_i;\hat\alpha_0)}{ \sum_{i=1}^n \one{\{V_{k,i}=1}\}  }.
\end{align*}
\noindent

\ignore{

Finally, under Theorem \ref{theo:modela} eq.~(\ref{eq:ecateexI}), 
we propose the following outcome regression estimator for $\ate(k)$,
\begin{align*}
    \hat{\ate}_{\text{OR}} &= \frac{\sum_{i=1}^n \mu_{\text{oc}}(1,w_i,e_i,1;\hat\beta_1)}{ n } - \frac{\sum_{i=1}^n \mu_{\text{all}}(0,w_i,e_i;\hat\alpha_0)}{ n   },
\end{align*}
\noindent
where $\mu_{\text{oc}}(1,w_i,e_i,1;\hat\beta_1)$ is learned among only concurrent and marginalized over the whole trial population by extrapolation. 
}

\paragraph{Large sample properties.} We derived the asymptotic properties of $\hat{\eate}_{\text{OR}}^{\text{oc}}$, and $\hat{\eate}_{\text{OR}}^{\text{all}}$, 
using the approach of M-estimation \citep[Chapter 7]{boos2013essential}. Under regularity conditions \citep[Section 7.2]{boos2013essential}, $\hat{\eate}_{\text{OR}}^{\text{oc}}$, $\hat{\eate}_{\text{OR}}^{\text{all}}$, 
are consistent and asymptotically Normal, with asymptotic variance derived in the appendix.

\subsection{An estimator based on parametric inverse probability weighting}
\label{sec:est_IPW}
Following standard procedures, we model the conditional probability of treatment assignment given $W$ and $E$ among only $V_k=1$ by using a logistic regression model,
\begin{align*}
   \E(\one{\{A=1}\} \mid W=w,E=e,V_k=1) &= \pi_{\text{oc}}(w,e,1;\eta) = \frac{\exp(\eta^T x)}{1+\exp(\eta^T x)}, 
\end{align*}
\noindent
where $x = (w, e, 1)$. We obtain an estimate of $\eta$ by using maximum likelihood estimation,  only among the concurrent controls $V_k=1$. We then propose,
$$\hat{\eate}_{\text{IPW}}^{\text{oc}} = \frac{\sum_{i=1}^n \gamma^1_i \one{\{V_{k,i}=1}\} y_i}{ \sum_{i=1}^n \gamma^1_i\  } -  \frac{\sum_{i=1}^n \gamma^0_i \one{\{V_{k,i}=1}\} y_i}{ \sum_{i=1}^n \gamma^0_i\  },$$
\noindent
where $\gamma^0_i = \one{\{A_{i}=0}\} / (1-\hat \pi_{\text{oc}})$, $\gamma^1_i = \one{\{A_{i}=1}\} / \hat \pi_{\text{oc}}$, and $\hat \pi_{\text{oc}} = \pi_{\text{oc}}(w,e,1;\hat \eta)$ for clarity. 
\paragraph{Large sample properties.} We derived the asymptotic properties of $\hat{\eate}_{\text{IPW}}^{\text{oc}}$ using the approach of M-estimation \citep[Chapter 7]{boos2013essential}. Under regularity conditions \citep[Section 7.2]{boos2013essential}, $\hat{\eate}_{\text{IPW}}^{\text{oc}}$, is consistent and asymptotically Normal, with asymptotic variance derived in the appendix. 

\subsection{Doubly robust estimators}
\label{sec:est_DR}
Doubly robust (DR) estimators for average treatment effects provide consistent estimates by combining outcome regression and IPW. To derive DR estimators of $\eate(k)$, we follow standard practice of constructing them based on efficient influence functions (EIF)s \citep{bickel1993efficient,fisher2021visually,hines2022demystifying,kennedy2022semiparametric}. Influence functions are a core component of classical statistical theory. They aid in constructing estimators with desirable properties such as double robustness, asymptotic normality, and fast rates of convergence. Additionally, they enable the incorporation of machine learning algorithms while preserving valid statistical inferences and providing insights into statistical efficiency, \textit{i.e.}, the best performance for estimating an estimand. We provide efficiency considerations of the proposed estimators in section \ref{sec:efficiency}. The next theorem provides these EIFs,

\begin{theorem}
\label{thm:eifs}
   The efficient influence function, $\varphi(Z,\eate(k))$, for $\eate(k)$ under Model (\ref{eq:npsem}), is equal to 
   \begin{multline}\label{eif1}
       \frac{\one{\{V_k=1\}}}{\P(V_k=1)}\bigg[\frac{2A-1}{\P(A\mid W, E, V_k=1)}\{Y-\E(Y\mid A, W, E, V_k=1)\}\\
       + \E(Y\mid A=1, W, E, V_k=1)-\E(Y\mid A=0, W, E, V_k=1) - \eate(k).\bigg]
   \end{multline}
    The efficient influence function, $\varphi(Z,\eate(k))$, for $\eate(k)$ under Model (\ref{eq:npsem}) and assuming \ref{ass:condex}, is equal to 
   \begin{multline} \label{eif2}
       \frac{\one{\{V_k=1\}}}{\P(V_k=1)}\bigg[\frac{A}{\P(A\mid V_k=1, W,E)}\{Y-\E(Y\mid A,W,E,V_k=1)\}
       \bigg]-\\
       \frac{1-A}{\P(A\mid W,E)}\frac{\P(V_k=1\mid E, W)}{\P(V_k=1)}\{Y-\E(Y\mid A, E, W)\}+\\
       \frac{\one{\{V_k=1\}}}{\P(V_k=1)}\Big[\E(Y\mid A=1,W,E,V_k=1)-\E(Y\mid A=0, W, E) \Big] -\eate(k) .
   \end{multline}
\end{theorem}  

\noindent
These influence functions suggest the following estimators,
 \begin{align*}
    \hat{\eate}_{\text{DR}}^{\text{oc}} &= \frac{1}{n} \sum_{i=1}^n \bigg( \frac{ \one{\{V_{k,i}=1\}} }{n^{-1}\sum_{i=1}^n \one{\{V_{k,i}=1\}}}  \bigg[ \frac{(2a_i-1)}{\hat \pi_{\text{oc}}(w_i,e_i,1)} \lbrace y_i - \hat \mu_{\text{oc}}(a_i,w_i,e_i,1) \rbrace  \\
    &+  \hat \mu_{\text{oc}}(1,w_i,e_i,1) - \hat \mu_{\text{oc}}(0,w_i,e_i,1) \bigg] \bigg) \\
    \hat{\eate}_{\text{DR}}^{\text{all}} &= \frac{1}{n} \sum_{i=1}^n \bigg( \frac{\one{\{V_{k,i}=1\}}}{n^{-1}\sum_{i=1}^n \one{\{V_{k,i}=1\}}}  \bigg[ \frac{ \one{\{A_{i}=1\}} }{\hat \pi_{\text{oc}}(w_i,e_i,1)} \lbrace y_i - \hat \mu_{\text{oc}}(1,w_i,e_i,1) \rbrace \bigg] \\
    &+ \frac{\one{\{A_{i}=0\}}}{1-\hat \pi_{\text{all}}(w_i,e_i)}\frac{\hat \nu(w_i,e_i)}{n^{-1}\sum_{i=1}^n \one{\{V_{k,i}=1\}}} \lbrace y_i - \hat \mu_{\text{all}}(0,w_i,e_i) \rbrace \\
    &+ \frac{\one{\{V_{k,i}=1\}}}{n^{-1}\sum_{i=1}^n \one{\{V_{k,i}=1\}}}  \bigg[ \hat \mu_{\text{oc}}(1,w_i,e_i,1) - \hat \mu_{\text{all}}(0,w_i,e_i) \bigg] \bigg),
  \end{align*}
\noindent
where, $\hat \pi_{\text{oc}}(w_i,e_i,1),\hat \mu_{\text{oc}}(1,w_i,e_i,1),\hat \mu_{\text{all}}(0,w_i,e_i)$, $\hat \pi_{\text{all}}(w_i,e_i)$, and $\hat \nu(w,e;\xi)$, can be estimated by using parametric and machine learning methods. 
Building on the results from the previous sections, we can leverage the linear regression models introduced earlier: $\mu_{\text{oc}}(0,w_i,e_i,1;\beta_0)$, and $\mu_{\text{all}}(0,w_i,e_i;\alpha_0)$  as outcome models. Similarly, the logistic regression models $\pi_{\text{all}}(w,e;\eta)$ and $\nu(w,e;\xi)$ introduced previously for $\P[A=1 \mid W=w, E=e]$ and $\P[V_k=1 \mid W=w, E=e]$, respectively, can also be employed.


\paragraph{Large sample properties.}  

Note that if $V_k$ is a deterministic function of $E$, and therefore $\P(V_k=1\mid E, W)=\one\{E>t\}$,
the two EIFs (\ref{eif1}) and (\ref{eif2}) are the same. In addition, the first influence function in Theorem \ref{thm:eifs} boils down to the standard influence function for the average treatment effect in the the $V_k=1$ population. Therefore, it inherits the standard analysis of the one-step estimator for average treatment effects as discussed in \cite[Section 4.1]{kennedy2021semiparametric}. Similar analysis can be conducted when $V_k$ is a stochastic function of $E$. In summary, if the outcome models are correctly specified, it can be shown that estimators of the form of $\hat{\eate}_{\text{DR}}^{\text{oc}}$ and  $\hat{\eate}_{\text{DR}}^{\text{all}}$ are root-n consistent, asymptotically normal with asymptotically valid 95\% confidence intervals given by the closed-form expressions $\hat{\eate}_{\text{DR}}^{\text{oc}} \pm 1.96 \sqrt{\hat{\text{var}} \lbrace \varphi(Z,\eate(k)) \rbrace/n}$ and $\hat{\eate}_{\text{DR}}^{\text{all}} \pm 1.96 \sqrt{\hat{\text{var}} \lbrace \varphi(Z,\eate(k)) \rbrace/n}$, respectively,  and efficient in the local asymptotic minimax sense. If the models are misspecified, the confidence intervals will be conservative \citep{kennedy2022semiparametric}. 

\paragraph{Double robustness.} Similarly, if $V_k$ is a deterministic function of $E$ 
the two estimators boils down to the standard DR estimator for the average treatment effect in the $V_k=1$ population, they also inherit the same double robust property. This means that if either the outcome regression model ($\mu_{\text{oc}}(0,w_i,e_i,1;\beta_0)$, $\mu_{\text{all}}(0,w_i,e_i;\alpha_0)$) or the treatment assignment model ($\pi_{\text{oc}}(w_i,e_i,1;\eta)$,  $\pi_{\text{all}}(w,e;\eta)$) is correctly specified (in a parametric sense), then the DR estimator is consistent, see section 4.2 of \cite{kennedy2022semiparametric} for details. Recall that our proposed estimators are doubly robust due to randomization \textit{i.e.} the treatment assignment mechanism is known by design.  We provide some empirical result of this property in our simulations in section \ref{sec:simu}. In addition, it can be shown that if $V_k$ is a stochastic function of $E$, the same double robustness property holds, provided the model for $V_k$ is correctly specified.

\section{Efficiency considerations}\label{sec:efficiency}


\paragraph{Estimators based on outcome regression.} As shown in the appendix, the influence function of the conditional expectation under control, $\varphi(Z_i,\hat \mu_0)$, depends on two components: the influence function of $\mu_0$ itself and that of the regression coefficients. Here, $\mu_0 = \E[Y(0) | V_k=1]$ depending on the estimand under study. Therefore, a more precise estimation of the regression coefficients (the variance of the estimated regression coefficient is inversely proportional to the sample size), translate to a more precise estimation of $\mu_0$ and consequently of $\hat{\eate}_{\text{OR}}^{\text{all}}$ compared to $\hat{\eate}_{\text{OR}}^{\text{oc}}$.

\paragraph{Doubly robust estimators.} As previously discussed, if $V_k$ is a deterministic function of $E$ 
the two EIFs presented in section \ref{sec:est_DR} are the same. In this case, efficiency gains come solely from better fitting of the regression $E(Y\mid A=0, W, E, V_k=1)$, which under assumption (\ref{ass:condex}) is equal to $E(Y\mid A=0, W, E, V_k=1)=E(Y\mid A=0, W, E)$ (because $V_k=\one\{E>t\}$). In this case it becomes purely about getting this regression right, and these efficiency gains do not show up in the first order analysis of the estimator. In addition, as aforementioned, efficiency gains can be obtained by leveraging prognostic variables as discussed in \cite{colantuoni2015leveraging,benkeser2021improving}. We show some empirical results in our simulations in section \ref{sec:simu}. Interestingly, efficiency gains can also be achieved when using doubly robust estimators under Model~(\ref{eq:npsem}) when $V_k$ is a non-deterministic function of $E$ as shown in our simulation setting. Under this scenario, we expect to see efficiency gains also when using an estimator based on inverse probability weighting solely.  


\ignore{
\begin{theorem}
   The efficient influence function for $\ate(k)$ int the non-parametric model that assumes (\ref{ass:extra}) is equal to 
          \begin{multline}
       \bigg[\frac{A}{\P(A\mid V=1, W,E)}\frac{\one{\{V=1\}}}{\P(V=1\mid E, W)} - \frac{1-A}{\P(A\mid V, W,E)}\bigg]\{Y-\E(Y\mid A, V, W,E)\}-\\
     \E(Y\mid A=1, V=1, W,E)-\E(Y\mid A=0, W, E)-\ate(k)
   \end{multline}
\end{theorem}
}





\section{Simulations}
\label{sec:simu}

In this section we evaluate the performance of the proposed estimators with respect to, bias squared, variance, mean square error, and coverage of the 95\% confidence interval, across levels of the percentage of concurrent controls, and model misspecification when estimating $\eate(k)$. 
We do not compare our proposed estimators with methods described in section  \ref{sec:relation}, because it is not clear if they target $\eate(k)$.

Finally, our simulations aim to showcase the theoretical properties previously discussed, rather than evaluating them under complex real-world scenarios. 

\subsection{Setup}
\label{simu_setup}

\paragraph{Aims} 
To evaluate the performance and gains in efficiency of our proposed estimators across levels of (1) percentage of concurrent controls (90\% to 10\%) and (2) model misspecification (correct outcome and treatment models; and misspecified outcome and correct treatment model) considering $V_k$ a deterministic function of $E$. In addition, we also evaluate efficiency gains by comparing the estimated variance of the outcome regression (Section \ref{sec:est_OR}) and doubly robust (Section \ref{sec:est_DR}) estimators that only use concurrent data compared with those that use all data considering $V_k$ both a deterministic and a stochastic function of $E$.

\paragraph{Data-generating mechanisms} We considered generating data from Model~(\ref{eq:npsem}). Specifically, we considered a sample size of $n=1,000$ and for each subject $i=1, \dots, n$, we simulated the following data \textcolor{black}{10,000 times}:
\begin{itemize}
    \item[] \textbf{Step 1.} the entry time $E  \sim \text{Norm}(0,1)$ and a baseline covariate $W = -\kappa_1 + 0.8 E + \text{Norm}(0,1)$, where $\kappa_1=n^{-1} \sum_{i=1}^n  0.8 E $;
    \item[] \textbf{Step 2.} an indicator whether treatment $k$ was available at time $E$, $V_k$ as a deterministic function of $E$ being less than a threshold describing the level of the percentage of concurrent controls;
    \item[] \textbf{Step 3.} a binary treatment $A \sim \text{Bernoulli}(\pi(W))$, where $\pi(W) = \left( 1 + \exp \left( - (-\kappa_2 + 0.8 W) \right)  \right)^{-1}$ and $\kappa_2=n^{-1} \sum_{i=1}^n  0.8 W $ when $V_k=1$ and $A=0$, otherwise  (participants for which treatment only control is available);
    \item[] \textbf{Step 4.} two counterfactual outcomes, $Y(0) = 0.8 W + 0.5 E + \text{Norm}(0,1)$, and $Y(k) = Y(0) + \Delta$, with $\Delta=0.8$, and the observed outcome $Y = AY(1) + (1-A)Y(0)$.
Since we consider a homogeneous treatment effect,  $\Delta=\eate(k)=0.8$ . 
\end{itemize}
\paragraph{Estimands} The estimand of interest is $\eate(k)$.
\paragraph{Methods} For each dataset across levels of percentage of concurrent controls, and misspecification we used the methods summarized in Table \ref{table_methods}. 
\paragraph{Performance metrics} Bias squared, variance, mean square error (MSE), and coverage of the 95\% confidence interval. In addition, we also considered the ratio of the estimated variances.
\paragraph{Scenarios} We considered levels of percentage of concurrent controls between 10\% and 90\% by 10\%. Misspecified models were set to only include an intercept -- not controlling for any covariates or entry time.

\begin{table}[]
\centering
\caption{Methods used in the estimation of $\eate(k)$. \label{table_methods}}
\begin{tabular}{lcc}
\hline
                                                                                                                                & \multicolumn{1}{c}{\textbf{Acronym}} \\  
\multicolumn{1}{c}{\textbf{Method}}                                                                                             & $\eate(k)$        \\ \hline
Outcome regression using only concurrent data, ($\hat{\eate}_{\text{OR}}^{\text{oc}}$, Section \ref{sec:est_OR})                & OR-oc            \\
Outcome regression using all data, ($\hat{\eate}_{\text{OR}}^{\text{all}}$, $\hat{\ate}_{\text{OR}}$, Section \ref{sec:est_OR}) & OR-ac            \\
Weighting using only concurrent data ($\hat{\eate}_{\text{IPW}}^{\text{oc}}$, Section \ref{sec:est_IPW})                        & IPW               \\
Doubly robust using only concurrent data ($\hat{\eate}_{\text{DR}}^{\text{oc}}$, Section \ref{sec:est_DR})                      & DR-oc             \\
Doubly robust  using all data ($\hat{\eate}_{\text{DR}}^{\text{all}}$, Section \ref{sec:est_DR})                                & DR-ac             \\ \hline
\end{tabular}
\end{table}

\ignore{

\begin{table}[]
\caption{Methods used in the estimation of $\eate$ and $\ate$. \label{table_methods}}
\begin{tabular}{lcc}
\hline
                                                                                                                                & \multicolumn{2}{c}{\textbf{Acronym}} \\ \cline{2-3} 
\multicolumn{1}{c}{\textbf{Method}}                                                                                             & $\eate(k)$        & $\ate(k)$        \\ \hline
Outcome regression using only concurrent data, ($\hat{\eate}_{\text{OR}}^{\text{oc}}$, Section \ref{sec:est_OR})                & OR-oc             & -                \\
Outcome regression using all data, ($\hat{\eate}_{\text{OR}}^{\text{all}}$, $\hat{\ate}_{\text{OR}}$, Section \ref{sec:est_OR}) & OR-ac             & OR-ad            \\
Weighting using only concurrent data ($\hat{\eate}_{\text{IPW}}^{\text{oc}}$, Section \ref{sec:est_IPW})    & IPW               & -                \\
Doubly robust using only concurrent data ($\hat{\eate}_{\text{DR}}^{\text{oc}}$, Section \ref{sec:est_DR})                                                             & DR-oc             & -                \\
Doubly robust  using all data ($\hat{\eate}_{\text{DR}}^{\text{all}}$, Section \ref{sec:est_DR})                                                                        & DR-ac             & -                \\ \hline
\end{tabular}
\end{table}

}

\subsection{Results}

\subsubsection{Bias, variance, MSE, and coverage}

\begin{figure}
\begin{center}
\includegraphics[scale=.55]{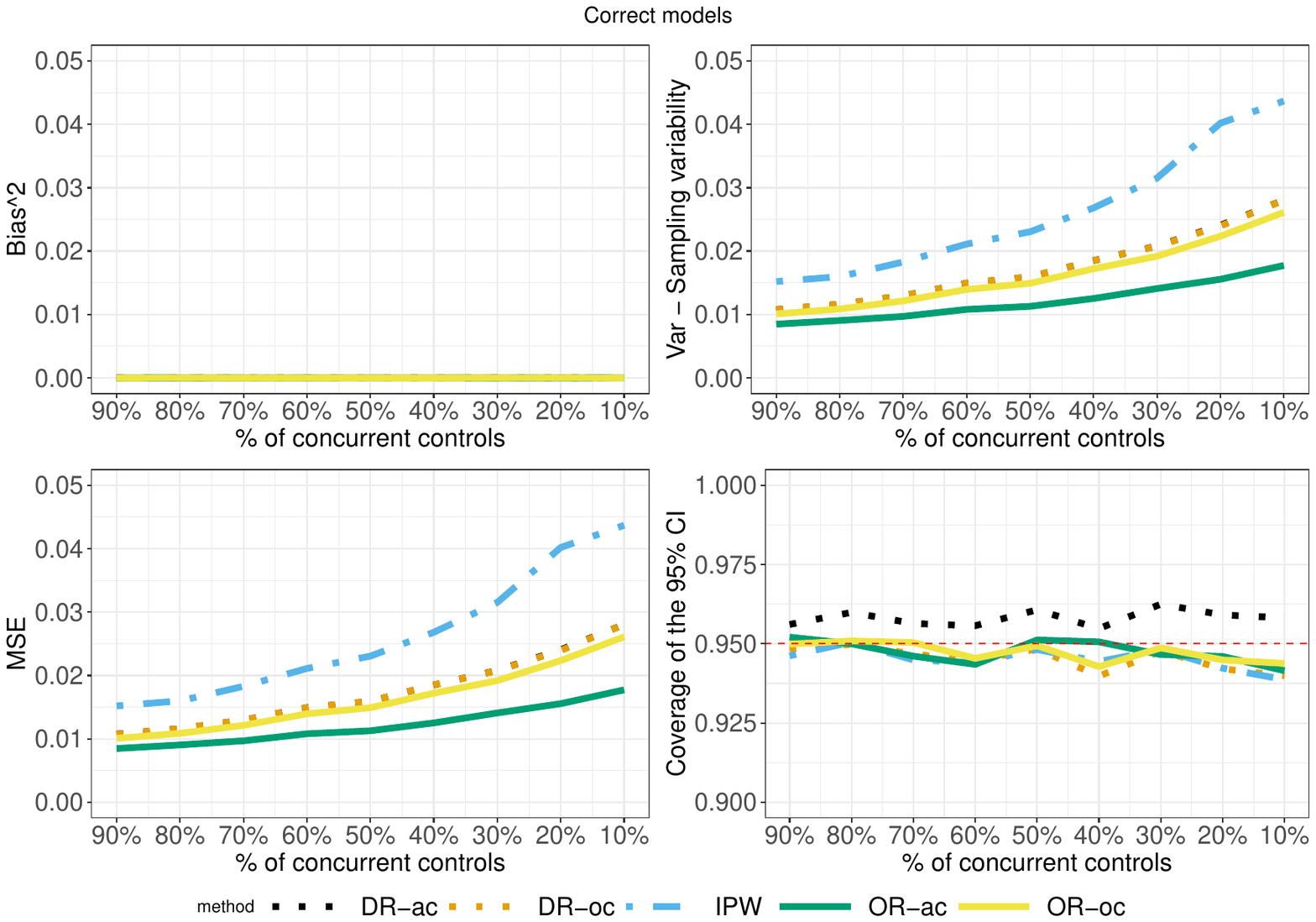}
\end{center}
\caption{ 
\label{fig_corr_corr} \textcolor{black}{Bias squared, variance, MSE and coverage of the 95\% confidence interval of DR-ac, DR-oc, IPW, OR-ac and OR-ac under correct models.  Note that, DR-ac and DR-oc overlap in terms of bias squared, sampling variability and MSE.}}
\end{figure}

\begin{figure}[h]
\begin{center}
\includegraphics[scale=.55]{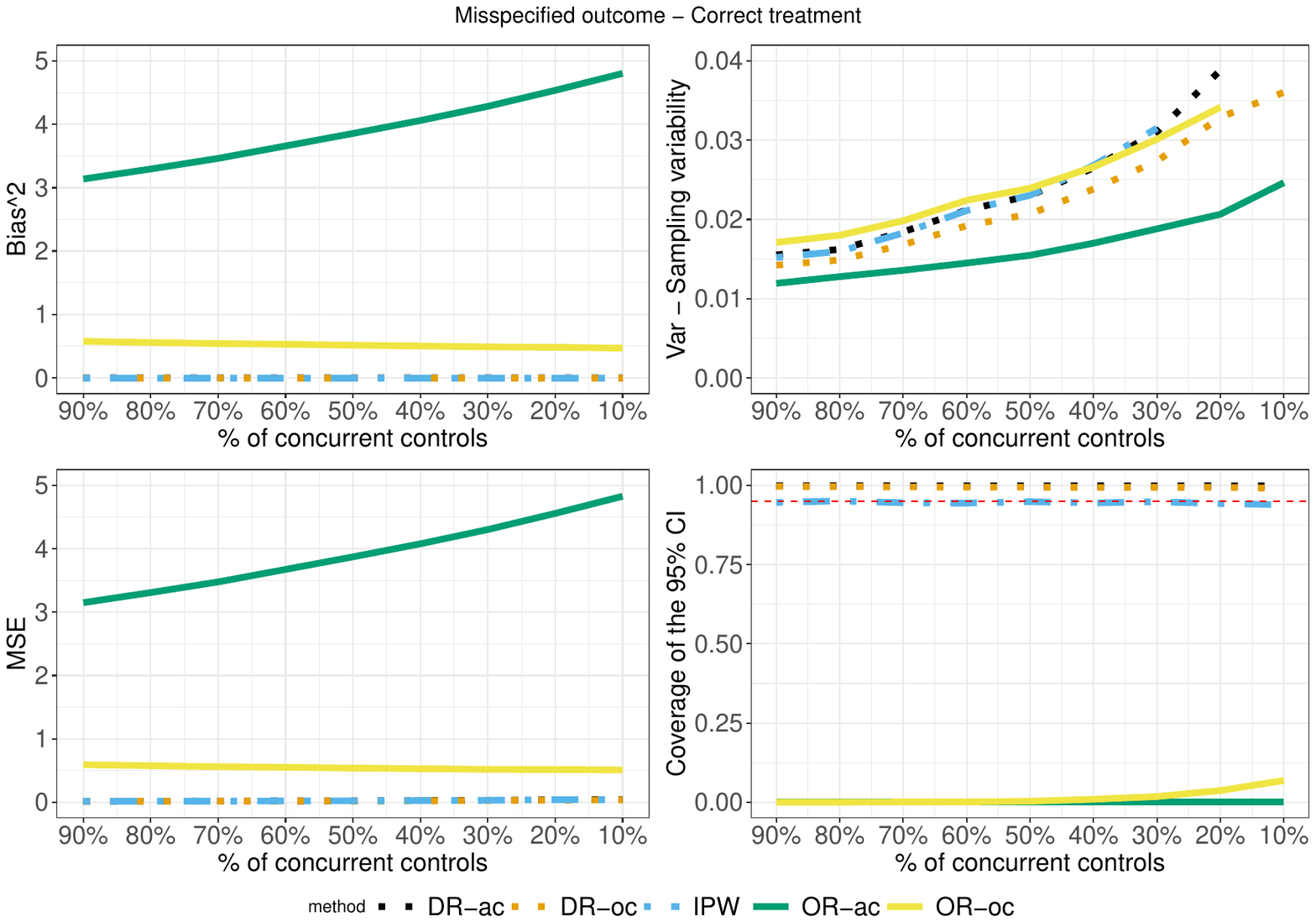}
\end{center}
\caption{ 
\label{fig_miss_corr} \textcolor{black}{Bias squared, variance, MSE and coverage of the 95\% confidence interval of DR-ac, DR-oc, IPW, OR-ac and OR-ac under misspecified outcome model and correct treatment assignment model. Note that, DR-ac and DR-oc overlap in terms of bias squared and MSE.}}
\end{figure}

Figure \ref{fig_corr_corr} and Figure \ref{fig_miss_corr}, show bias squared, variance, MSE and coverage of the 95\% confidence intervals in estimating $\eate(k)$ across percentage of concurrent controls when models are correct and when only the treatment model is correctly specified, respectively.  When both the outcome and the treatment models are correct (Figure \ref{fig_corr_corr}), bias squared is negligible across levels of concurrent controls for all methods. Variance is shown to increase with decreasing levels of concurrent controls across all methods, with OR-ac being smaller than OR-oc, suggesting a gain in efficiency (more on this in the next section). Similar behavior can be seen for the MSE. Finally, all methods achieve desirable coverage levels.  When the outcome model is misspecified (Figure \ref{fig_miss_corr}), both estimators based on outcome regression show bias while maintaining a relatively small variance. MSE is consequently dominated by bias. DR estimators and the IPW estimator maintain negligible levels of bias and relatively small variance. 

\subsubsection{Efficiency gains}

\begin{figure}
\begin{center}
\includegraphics[scale=.55]{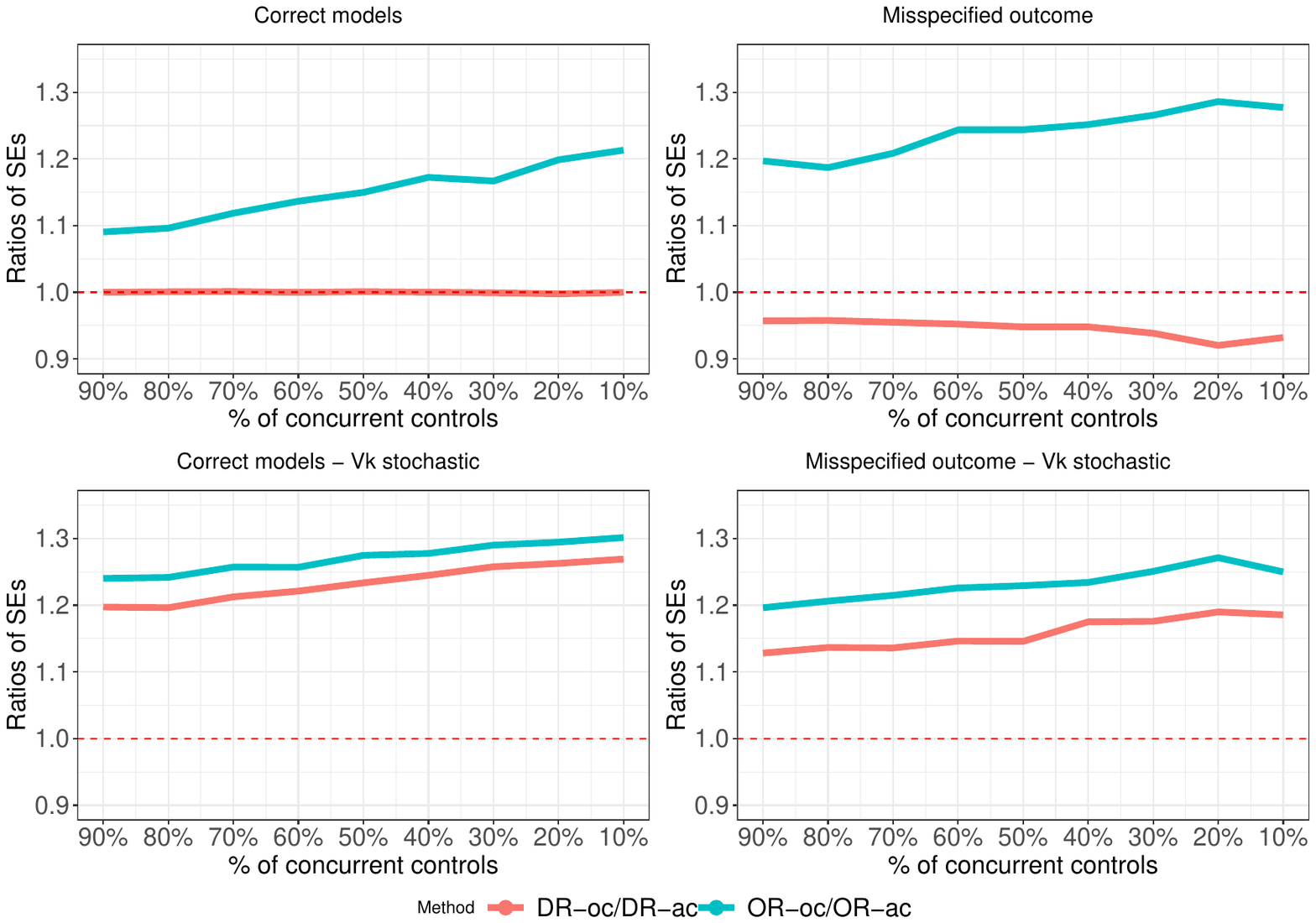}
\end{center}
\caption{ 
\label{fig_ratio} \textcolor{black}{Ratio of standard errors DR-oc/DR-ac and OR-oc/OR-ac across model misspecifications considering $V_k$ a deterministic function of $E$ (top panels) and not (bottom panels). A ratio greater than 1 means efficiency gains.}}
\end{figure}

Figure \ref{fig_ratio} shows the ratio of the estimated standard errors of DR-oc over DC-ac and OR-oc over OR-ac across levels of concurrent controls and misspecification. 

The top panels of Figure \ref{fig_ratio} follow Model~(\ref{eq:npsem}) where $V_k$ is a deterministic function of $E$.
As discussed in Section \ref{sec:efficiency}, estimators based on outcome regression that use all controls seem to have a gain in efficiency, while DR estimators did not under correct models. 

The bottom panels of Figure \ref{fig_ratio} follow Model~(\ref{eq:npsem}) where $V_k$ is a not a deterministic function of $E$. Specifically, we generated $V_k \sim \text{Bernoulli}(\pi(E))$, where $\pi(E) = \left( 1 + \exp \left( - (-\kappa_4 + 0.5 E) \right)  \right)^{-1}$ and $\kappa_4= \kappa_3 + n^{-1} \sum_{i=1}^n  0.5 E $, and $\kappa_3$ is the threshold discussed in Step 2 above. Under correct models (including the one for $V_k$), the entry time $E$ is a prognostic variable of $V_k$ and therefore improves efficiency compared to using only concurrent controls (Bottom panels of \ref{fig_ratio}; DR-oc/DR-ac). These results suggest that efficiency gains can be obtained when using a doubly robust estimator under Model~(\ref{eq:npsem}) when $V_k$ is not a deterministic function of $E$ and the model for $V_k$ is correct.  

\paragraph{Summary of results.} Methods based on outcome regression improve efficiency when using non-concurrent controls. However, they introduce bias when misspecified. In contrast, doubly robust estimators provide consistent estimates with relatively small variance when either the treatment or outcome model is correctly specified. In addition, doubly robust estimators have the potential to additionally improve efficiency when $V_k$ is not a deterministic function of $E$ under Model~(\ref{eq:npsem}). \textcolor{black}{The relatively small efficiency gains between 90\% and 10\% of concurrent controls, as shown in the top left panel of Figure \ref{fig_ratio}, can be attributed to plotting the ratio of standard errors rather than the ratio of variances, which range from 1.20 to 1.50, respectively. Additionally, the efficiency gain is primarily due to improved estimation of the expected response under $A = 0$. When focusing solely on efficiency gains among controls, the variance improves from 1.29 to 1.80, while the standard error increases from 1.10 to 1.35, for 90\% and 10\% of concurrent controls, respectively. Similar results were obtained when considering a smaller sample size of n=100 (results in the appendix).}

\section{Practical considerations}
\label{sec:pract_guide}

\paragraph{What estimand should we target?} In this paper, we introduce $\eate(k)$, whereas it is not clear what estimand the current related literature targets. 
Given the more stringent and untestable assumptions needed to identify $\ate(k)$, we suggest targeting $\eate(k)$. 
Note that $\ate(k)$ and $\eate(k)$ will coincide under the assumption of homogeneous treatment effect. While this is true in theory, we would expect it not to hold in practical settings, leading to different results as shown in our case study in the next section. 

\paragraph{Should we pool concurrent and non-concurrent controls? Testing assumption~\ref{ass:condex}.} To evaluate assumption~\ref{ass:condex} under Model~(\ref{eq:npsem}), we propose using the method introduced by  \cite{luedtke2019omnibus}. Specifically, following the notation of the original paper, we suggest to set $R_P(o) \triangleq \E(Y\mid A, W, E, V_k=1)\ - 
 \E(Y\mid A=0, W, E)$ and $S_P(o) \equiv 0$, where $o$ is the observed data, and $R_P$ and $S_p$ are elements of the space of univariate bounded real-valued measurable functions defined on the support of the distribution $P$ \citep{luedtke2019omnibus}.  We show an application of this testing procedure in our case study. Note that if $V_k$ is a deterministic function of $E$ as in our case study, this test is a test on the statistical models, \textit{i.e.}, if the projections on the linear model are different, and not a test on the true expectations. In other words, it is a test to evaluate model misspecifications and therefore guide analysts on the choice to either use or not non-concurrent controls.


\paragraph{Should we leverage prognostic baseline variables for additional precision?} Recent literature suggests that incorporating baseline prognostic variables can improve the precision of estimates \citep{colantuoni2015leveraging}. We propose following this approach by appropriately controlling for these variables in the analysis. This may explain the increased precision observed with DR-oc and OR-oc estimators compared to the naive estimator in our case study (presented in the next section), despite being computed within the concurrent population only.

\paragraph{What estimator should we use?}


Under Model~(\ref{eq:npsem}) and assuming $V_k$ is a deterministic function of $E$, to estimate $\eate(k)$, we recommend using $\hat{\eate}_{\text{DR}}^{\text{oc}}$, the doubly robust estimator using only concurrent data (Section \ref{sec:est_DR}). We recommend this estimator because: 1) it has the same efficiency as the DR estimator that uses non-concurrent controls 
2) it does not require any additional assumptions, thus better aligning with the FDA recommendations \citep[Section A.5.1]{food2021e9}; 3) it accommodates covariates that, if prognostic, can be leveraged to improve efficiency \citep{colantuoni2015leveraging}; and 4) it is doubly robust, meaning that it is consistent when either the treatment assignment model or the outcome model is correctly specified, a property we obtain “for free” in platform trials due to randomization. 
Assuming $V_k$ is a stochastic function of $E$ and if an analyst chooses to leverage non-concurrent controls, then the DR-ac estimator is recommended.

\color{black}

\paragraph{\textcolor{black}{Multiple treatment arms in addition to a shared arm.}}
While Model~(\ref{eq:npsem}) and our proposed estimators apply to multiple treatment arms, for simplicity, the simulation section focuses on a single treatment arm compared to a shared control. Additional results for scenarios with two treatment arms and a shared control are provided in the Appendix. Under this expanded scenario, it is important to note that multiple concurrent average treatment effects can be considered. For example, in our simulation setting, we considered the following for treatment $A$: $\E[Y(A) - Y(0) | V_A=1]$, $\E[Y(A) - Y(0) | V_A=1, V_B=0]$, and $\E[Y(A) - Y(0) | V_A=1, V_B=1]$. Similarly, for treatment $B$, we evaluated $\E[Y(B) - Y(0) | V_B=1]$, $\E[Y(B) - Y(0) | V_A=0, V_B=1]$, and $\E[Y(B) - Y(0) | V_A=1, V_B=1]$. In our simulations, we considered approximately 60\% for $V_A=1$, 40\% for $V_B=1$, 50\% for $V_A=1,V_B=0$, 30\% for $V_A=0,V_B=1$ and 10\% for $V_A=V_B=1$. Consistent with our simulation results, efficiency gains were observed only when comparing OR-ac with OR-oc under the condition $V_A=V_B=1$, for both treatments $A$ and $B$. For other scenarios and estimators, the standard errors were comparable, with no notable efficiency improvements.

\color{black}

\paragraph{Sample size calculation for a prospective trial with non-concurrent controls.} Our theoretical and methodological results suggest an efficiency gain when including non-concurrent control with estimators based on regression models. While these results are promising, we suggest to conduct standard sample size calculation as if the non-concurrent control data will not be available. At the analysis stage, precision can then be improved by using non-concurrent control data as previously described with the caveat that the outcome model must be correctly specified. 

\paragraph{Multiple comparisons.} Our proposed methods enable the use of standard type I error control procedures, such as Bonferroni or Benjamini-Hochberg corrections, due to the validity of 95\% confidence intervals, test statistics, and p-values (demonstrated in previous sections). 
\textcolor{black}{This allows for direct application of these corrections in platform trials with, for instance, multiple primary endpoints.}


\paragraph{Summary of practical guidelines.} Target $\eate(k)$ as the estimand of interest. Use DR-oc to obtain consistent estimates of $\eate(k)$. To improve efficiency, focus on prognostic baseline covariates rather than relying on non-concurrent controls. If leveraging non-concurrent controls is of interest, use DR-ac. Finally, conduct sample size calculations without considering non-concurrent controls and conduct standard multiple comparisons adjustments.



\section{The Adaptive COVID-19 Treatment Platform Trial}
\label{case_study}

In this section, we apply our proposed estimators to estimate $\eate$ using data from the Adaptive COVID-19 Treatment Trial (ACTT) \citep{kalil2021baricitinib}. 
This was a platform trial that investigated treatments for hospitalized adult patients with COVID-19 pneumonia. The trial comprised multiple stages, as illustrated in Figure \ref{fig:nonconc}. The initial phase, ACTT-1, involved the assessment of the effectiveness of remdesivir alone compared to placebo. Subsequently, in the second stage (ACTT-2), the placebo was phased out, and a novel treatment, combining remdesivir with baricitinib, was introduced. Simultaneously, participants were randomized to receive either remdesivir alone or the combination therapy of remdesivir and baricitinib. Data were accessed using the NIAID Clinical Trials Data Repository (\hyperref[https://data.niaid.nih.gov/]{https://data.niaid.nih.gov/}). We have a Data User Agreement in place for its use.

\paragraph{Study population and endpoint of interest.} We considered the combined participants of ACCT-1 and ACTT-2 as our study population. We followed the inclusion and exclusion criteria of the original study. The final study population was comprised of 1,379 participants, 541 from ACTT-1 and 1,033 from ACTT-2. We considered the time to recovery in days as our enpoint of interest. 

\begin{figure}
\begin{center}
\includegraphics[scale=0.55]{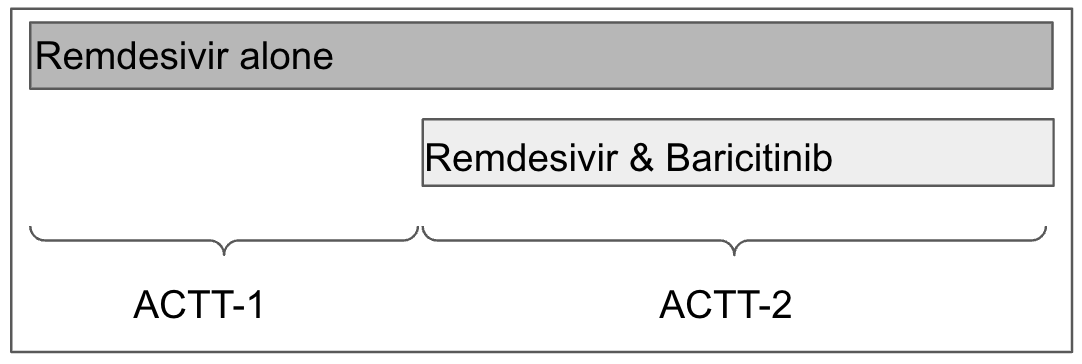}
\end{center}
\caption{ 
\label{fig:case-study} Case-study summary: We considered the two treatment arms: remdesivir alone (the shared control arm) and remdesivir plus baricitinib. We calculated $\eate(1)$ for the ACTT-2 population, where 1 represents the remdesivir plus baricitinib arm and 0 represents the remdesivir alone arm.}
\end{figure}

\paragraph{Treatments under study and targeted causal estimand.}
We considered two treatment arms: remdesivir alone (which served as the shared control arm) and remdesivir plus baricitinib. Our target estimand was $\eate(1)$, where $1$ represents the remdesivir plus baricitinib arm and $0$ represents the remdesivir alone shared arm. 
\textcolor{black}{While the original ACTT-1 study included an interim analysis that dropped the placebo arm, our case study focuses solely on the remdesivir and remdesivir-plus-baricitinib arms. Figure \ref{fig:case-study} illustrates the design of our case study.}

\paragraph{Baseline covariates.} We consider the following baseline covariates: age, sex assigned at birth (female, male), race (White, Black, Asian, Other: American Indian or Alaska Native, Native Hawaiian or Other Pacific Islander, Multiple), ethnicity (Hispanic or Latino, Not Hispanic or Latino), BMI, geographic region of study site (Asia, Europe, North America), disease severity stratum (mild, severe), and having any of these comorbidities: duration of symptoms, hypertension, coronary artery disease, congestive heart disease, chronic oxygen requirement, chronic respiratory disease, chronic liver disease, chronic kidney disease, diabetes type I, diabetes type II, obesity, cancer, immune deficiency, and asthma, in addition to the entry time which we normalized to be between 0 and 1. 


\paragraph{Models setup.} We computed 
OR-oc, OR-ac, OR-ad, IPW, DR-oc and DR-ac (Table \ref{table_methods}) by using linear and logistic regression models. We computed the naive estimator by taking the average difference in the endpoint between the two arms among only concurrent participants. Variances were obtained by using the sandwich estimator (for naive, OR-oc, OR-ac, IPW) and by taking the variance of the efficient influence function (for DR-oc and DR-ac). Wald 95\% confidence intervals and Wald tests were constructed. 

\paragraph{Results.} Table \ref{table_ACTT} shows the point estimate for $\eate(1)$, standard errors, 95\% confidence intervals and p-values. The naive estimate of  $\eate(1)$,  resulted in a value of -1.33 with a standard error of 0.58. This suggests that baricitinib plus remdesivir was superior to remdesivir alone in reducing recovery time as in the original trial \citep{kalil2021baricitinib}. OR-oc, IPW, DR-oc and DR-ac improved precision while maintaining a similar point estimate. OR-ac improved precision the most (around 28\% improvement compared with the naive estimator), however, it resulted in a different point estimate, -0.75 which led to a non-significant result. 
This suggests that the outcome model used to obtain OR-ac might be misspecified and therefore that assumption \ref{ass:condex} does not hold.  
 Using an omnibus test as described in our practical guidelines in section \ref{sec:pract_guide} we obtained a p-value $\leq 0.001$ using both the variance and the eigenvalue approach \citep{luedtke2019omnibus}, thus supporting rejecting assumption \ref{ass:condex} and, therefore, suggests not using non-concurrent controls. In contrast, doubly robust estimators improved precision while maintaining a similar point estimate as the naive estimator.  We believe the improved precision observed in OR-oc, IPW, and DR-oc compared to the naive estimator stems from appropriately adjusting for baseline variables, as discussed in \cite{colantuoni2015leveraging}, and in our previous section.
Note that a conditional analysis, targeting $\ate(1)$, using a standard linear regression model regressing the outcome in the full population on treatment arms, entry time and baseline covariates led to a non-significant point estimate of -0.46 (standard error equal to 0.52). 


\begin{table}[]
\caption{Estimated $\eate(1)$ using the ACTT data. \label{table_ACTT}}
\centering
\begin{tabular}{lcccccc}
\hline
\multicolumn{1}{c}{\textbf{Method}} & \textbf{$\hat \eate(1)$} & \textbf{SE} & \textbf{95\% CI} & \textbf{p-value} & \textbf{Ratio} \\ \hline
\textbf{OR-oc}                      & -1.29                         & 0.47                  & (-2.21;-0.37)       & $<$0.01           & 1.22        \\
\textbf{OR-ac}                      & -0.75                        & 0.45                  & (-1.63;0.13)        & 0.10           & 1.28         \\
\textbf{IPW}                        & -1.28                         & 0.47                  & (-2.20;-0.36)       & $<$0.01           & 1.22         \\
\textbf{DR-oc}                      & -1.30                        & 0.47                  & (-2.22;-0.38)       & $<$0.01           & 1.22         \\
\textbf{DR-ac}                      & -1.30                         & 0.47                  & (-2.22;-0.38)       & $<$0.01           & 1.21         \\
\textbf{naive}                      & -1.33                       & 0.58                  & (-2.47;-0.19)       & 0.02           & 1.00         \\
\hline
\end{tabular}
\end{table}

\section{Conclusion}

In this paper, we introduced identification results and estimation techniques to identify and estimate concurrent average treatment effects $\eate(k)$ in the presence of non-concurrent control in platform trials. We argue that identifying and estimating $\ate(k)$ relies on an extrapolation assumption that is both untestable and often too stringent, particularly in the context of platform trials, where multiple, potentially novel treatments or interventions are being evaluated and the outcome mechanism is poorly understood. Therefore, we advocate focusing primarily on $\eate(k)$, where assumptions can be tested. 
By focusing on $\eate(k)$ rather than $\ate(k)$, we also open the door to leveraging non-parametric models based on machine and deep learning techniques for learning outcome and treatment assignment mechanisms under the proposed doubly robust estimators \citep{kennedy2022semiparametric,diaz2020machine,hirshberg2021augmented}. In fact, while these methods can capture complex data relationships, potentially mitigating model misspecification, they may not be suitable for extrapolation.
Furthermore, our proposed doubly robust estimator accommodates Bayesian techniques while retaining valid frequentist properties, as demonstrated in \citep{shin2023improved,antonelli2022causal}. 

A key takeaway of this paper is to target $\eate(k)$ in platform trials, and, under Model~(\ref{eq:npsem}), to use a doubly robust estimator only among concurrent units, prioritizing the identification of strong prognostic baseline variables rather than relying on non-concurrent controls, especially when $V_k$ is a deterministic function of $E$.


In this paper, we presented results that can be used for continuous and binary endpoints. Estimators can be constructed for time-to-event endpoints under the non-parametric causal model introduced in eq. (\ref{eq:npsem}). 

Finally, in this paper, we demonstrate results assuming a structural equation model where treatment assignment may depend on baseline covariates; however, similar identification and estimation results can be obtained without baseline covariates.

\bibliographystyle{agsm}
\bibliography{refs,refs_a_exp, nonconc}




\bibliographystyle{agsm}
\bibliography{refs,refs_a_exp, nonconc}

\section*{Acknowledgments}
The author gratefully acknowledge Kelly Van Lancker for valuable comments on this paper.

\subsection*{Financial disclosure}

This article is based upon work supported by the National Science Foundation under Grant No 2306556.

\subsection*{Conflict of interest}

The authors declare no potential conflict of interests.

\newpage
\setcounter{page}{1}
\setcounter{assumption}{0}
\setcounter{remark}{0}
\begin{center}
{\large\bf SUPPLEMENTARY MATERIAL \\}
{\large\bf Identification and estimation of causal effects using non-concurrent controls in platform trials \\}
\author{Michele Santacatterina,
    Federico Macchiavelli Giron, 
    Xinyi Zhang, and     Iv\'an D\'iaz \\
    Division of Biostatistics, Department of Population Health, \\ New York University School of Medicine, \\ 
    New York, NY, 10016}
\end{center}

\begin{description}


\item[Proofs and M-estimation details:] Theorem \ref{theo:modela}, \ref{theo:modela2}, and \ref{thm:eifs} and M-estimation details for estimators based on outcome regression and parametric weighting.

\end{description}

\newpage

\setcounter{page}{1}
\setcounter{assumption}{0}
\setcounter{remark}{0}

\subsection*{Proof of Theorem 1}

    \begin{figure}[h]
        \centering
        \input{dag-b2.tex}
        \caption{DAG associated to the structural equation model in equation    (\ref{eq:npsem2}).}     \label{fig:dag-b2}
    \end{figure}
    \begin{figure}[h]
        \centering
        \begin{align}
          E_i&=f_E(U_{E,i}),\notag\\
          W_i&=f_W(E_i, U_{W,i}),\notag\\
          V_{k,i}&=f_{V_k}(E_i, U_V),\label{eq:npsem2}\\
          A_i&=f_A(V_i, W_i, U_{A,i}),\notag\\
          Y_i&=f_{Y}(A_i, W_i, E_i, U_{Y,i}).\notag
        \end{align}
    \end{figure}

\vspace{0.2in}

\noindent 
Since we are interested in the effect of $A$ on $Y$, and in using non-concurrent controls, $V_k=0$, we study paths from $A$ to $Y$ and then apply d-separation (we avoided augmenting the graph with $Y(k)$ and also not added the exogenous variables for clarity). We start by studying paths from $A$ to $Y$.
\begin{align*}
    &A \leftarrow V_k \leftarrow E \rightarrow Y && \lbrace V_k \rbrace; \lbrace E \rbrace ; \lbrace V_k,E \rbrace \\
    &A \leftarrow V_k \leftarrow E \rightarrow W \rightarrow Y && \lbrace V_k \rbrace; \lbrace E \rbrace; \lbrace W \rbrace; \lbrace V_k,E \rbrace; \lbrace V_k,W \rbrace; \lbrace E,W \rbrace; \lbrace V_k,W,E \rbrace \\
    &A \leftarrow W  \rightarrow Y && \lbrace W \rbrace \\
    &A \leftarrow W \leftarrow E  \rightarrow Y && \lbrace W \rbrace; \lbrace E \rbrace; \lbrace W,E \rbrace .
\end{align*}
\noindent
By applying d-separation, the set $\lbrace V_k,W,E \rbrace$ conditionally block the path from $A$ to $Y$. This leads to the following assumptions:

\begin{assumption}[weak A-ignorability]\label{ass:wA-ignorability_2} Let $a=0,\ldots,K$. Assume \\
    $\E[Y(k) | W=w,E=e,V_k=v] = \E[Y(k) | A=k,W=w,E=e,V_k=v]$.
\end{assumption}






\begin{assumption}[Consistency]\label{ass:cons_2} Assume \\
$\P(Y(k) | A=k, W=w, E=e, V_k=v) = \P(Y | A=k, W=w, E=e, V_k=v)$.
\end{assumption}

\begin{assumption}[Positivity of treatment assignment mechanism among
  concurrent units]\label{ass:posa_2}
  Assume \\$\P(A=k\mid W=w, E=e, V_k=1)>0$ for all $w$ and $e$ s.t. $V_k=1$.
\end{assumption}

\begin{assumption}[Positivity of shared arm assignment mechanism among all controls]\label{ass:posa0_2}
  Assume \\$\P(A=0\mid W=w, E=e)>0$ for all $w$ and $e$.
\end{assumption}


\begin{assumption}[Pooling concurrent and non-concurrent controls]\label{ass:condex_2}Assume  \\
$\E(Y\mid A=0, W=w, E=e, V_k=1) = 
  \E(Y\mid A=0, W=w, E=e)$ for all $e$ s.t. $V_k=1$.
  \end{assumption} 
\begin{assumption}[Conditional exchangeability of outcome mechanism
  among the treated]\label{ass:extra_2} Assume \\
  $\E(Y \mid A=k, W=w, E=e, V_k=1)= \E(Y\mid A=k, W=w, E=e)$ for all $e$.\\
\end{assumption}

\subsection*{Identification of concurrent ATE}

Recall that 

\begin{definition}[Conditional and marginal average treatment effect of treatment
  arm $k$ compared to control among concurrent population]
  \begin{align*}
    \ecate(k,w,e) &= \E[Y(k) - Y(0)\mid V_k=1, W=w, E=e]\\
    \eate(k) &= \E[\ecate(k,W,E)\mid V_k=1].
  \end{align*}
\end{definition}

\subsubsection*{Identification based on the G-formula.}
\label{sec:ident_ccate_g}
\begin{proof}
We start by showing it for treatment $k$. We refer to $W=w, E=e$ as $W,E$ for clarity and (IE) as iterated expectation. 
\begin{align*}
    &\E(Y(k) \mid V_k=1) \\ &= \E( \E(Y(k) | V_k=1, W, E) \mid V_k=1)  &&\text{by (IE)} \\
    &= \E( \E(Y(k) | A=k, V_k=1, W, E) \mid V_k=1)  &&\text{by (\ref{ass:wA-ignorability_2})} \\
    &= \E( \E(Y | A=k, V_k=1, W, E) \mid V_k=1)  &&\text{by (\ref{ass:cons_2},\ref{ass:posa_2})} \\
    &= \frac{ 1}{\P(V_k = 1)} \E(\mathds{1}[V_k=1]\E(Y | A=k, V_k=1, W, E) ) 
\end{align*}

\noindent
We now show the proof under treatment 0.

\begin{align*}
    &\E(Y(0) \mid V_k=1) \\ &= \E( \E(Y(0) | V_k=1, W, E) \mid V_k=1)  &&\text{by (IE)} \\
    &= \E( \E(Y(0) | A=0, V_k=1, W, E) \mid V_k=1)  &&\text{by (\ref{ass:wA-ignorability_2})} \\
    &= \E( \E(Y | A=0, V_k=1, W, E) \mid V_k=1)  &&\text{by (\ref{ass:cons_2},\ref{ass:posa0_2})} \\
    &= \E( \E(Y | A=0, W, E) \mid V_k=1)  &&\text{by (\ref{ass:condex_2})} \\
    &= \frac{ 1}{\P(V_k = 1)} \E( \mathds{1}[V_k=1] \E(Y | A=0, W, E))  
\end{align*}

\noindent
Consequently, under (\ref{ass:wA-ignorability_2})-(\ref{ass:posa0_2}), 
$\ecate(k,w,e)$ is non-parametrically identified as \label{stat:1}
\begin{equation}
    \E(Y\mid A=k, V_k=1,W,E) - \E(Y\mid A=0, V_k=1, W,E).
\end{equation}
\noindent
In addition, under (\ref{ass:wA-ignorability_2})-(\ref{ass:condex_2}),
$\ecate(k,w,e)$ is identified as
\begin{equation}
    \label{eq:ecateexI_2}
    \E(Y\mid A=k, V_k=1, W,E) - \E(Y\mid A=0, W, E).    
\end{equation}

\end{proof}


\ignore{

\begin{proof}
We start by showing it for treatment $k$.
    \begin{align*}
        \E(Y(k) \mid V_k=1, W=w,E=e) &= \E(Y(k) \mid A=k, V_k=1, W=w,E=e) &&\text{by (\ref{ass:wA-ignorability})} \\
        &= \E(Y \mid A=k, V_k=1, W=w,E=e) &&\text{by (\ref{ass:cons_2})}
    \end{align*}
\noindent     
We now show it for treatment $0$.
    \begin{align*}
        \E(Y(0) \mid V_k=1, W=w,E=e) &= \E(Y(0) \mid A=0, V_k=1, W=w,E=e) &&\text{by (\ref{ass:wA-ignorability_2})} \\
        &= \E(Y \mid A=0, V_k=1, W=w,E=e) &&\text{by (\ref{ass:cons_2})}
    \end{align*}
\end{proof}

\noindent
In the manuscript we state that, under \ref{ass:condex_2}, $\ecate(k,w,e)$ is 
identified as \label{stat:2}
  \begin{equation}
    \label{eq:ecateexI}
    \E(Y\mid A=k, V_k=1, W=w,E=e) - \E(Y\mid A=0, W=w, E=e).    
  \end{equation}

\begin{proof}
We only show it for treatment $0$ because the treatment $k$ follows the proof above.
  \begin{align*}
        \E(Y(0) \mid V_k=1, W=w,E=e) &= \E(Y(0) \mid A=0, V_k=1, W=w,E=e) &&\text{by (\ref{ass:wA-ignorability_2})} \\
        &= \E(Y \mid A=0, V_k=1, W=w,E=e) &&\text{by (\ref{ass:cons_2})} \\
        &= \E(Y \mid A=0, W=w,E=e) &&\text{by (\ref{ass:condex_2})} 
    \end{align*}
\end{proof}

}

\subsubsection*{Identification based on weighting.}




\begin{proof}
We start by showing it for treatment $k$. We refer to $W=w, E=e$ as $W,E$ for clarity, and (IE) as iterated expectation. 
\begin{align*}
    &\E(Y(k) \mid V_k=1) \\ &= \E( \E(Y(k) | V_k=1, W, E) \mid V_k=1)  &&\text{by (IE)} \\
    &= \E( \E(Y(k) | A=k, V_k=1, W, E) \mid V_k=1)  &&\text{by (\ref{ass:wA-ignorability_2})} \\
    &= \E( \E(Y | A=k, V_k=1, W, E) \mid V_k=1)  &&\text{by (\ref{ass:cons_2})} \\
    &= \E \left( \E \left( \frac{\mathds{1}[A=k,V_k=1] Y}{\P(A=k | V_k=1, W, E)} | W, E \right) \mid V_k=1 \right)  &&\text{by (\ref{ass:posa_2})} \\
    &= \frac{1}{\P(V_k = 1)} \E \left( \mathds{1}[V_k=1] \E \left(  \frac{\mathds{1}[A=k,V_k=1]Y}{\P(A=k | V_k=1, W, E)} | W, E \right) \right)   \\
    &= \frac{1}{\P(V_k = 1)} \E \left( \E \left(  \frac{\mathds{1}[A=k,V_k=1]Y}{\P(A=k | V_k=1, W, E)} | W, E \right) \right)   \\
    &= \frac{1}{\P(V_k = 1)} \E \left( \E \left(  \frac{\mathds{1}[A=k]Y\P(V_k=1|W,E)}{\P(A=k | V_k=1, W, E)} | W, E \right) \right)  &&\text{by (IE)} \\
    &= \frac{1}{\P(V_k = 1)} \E \left(  \frac{\mathds{1}[A=k]Y\P(V_k=1|W,E)}{\P(A=k | V_k=1, W, E)} \right)   \\
\end{align*}

\noindent
Note that if $V_k$ is deterministic, then $\P(V_k=1|W,E) = \mathds{1}[E>t]  = \mathds{1}[V_k=1]$ and therefore 

\begin{align*}
    \frac{1}{\P(V_k = 1)} \E \left(  \frac{\mathds{1}[A=k]Y\P(V_k=1|W,E)}{\P(A=k | V_k=1, W, E)} \right)   
    = \frac{1}{\P(V_k = 1)} \E \left(  \frac{\mathds{1}[A=k]Y\mathds{1}[V_k=1]}{\P(A=k | V_k=1, W, E)} \right). 
\end{align*}

\noindent
We now show the proof under treatment 0.

\begin{align*}
    &\E(Y(0) \mid V_k=1) \\ &= \E( \E(Y(0) | V_k=1, W, E) \mid V_k=1)  &&\text{by (IE)} \\
    &= \E( \E(Y(0) | A=0, V_k=1, W, E) \mid V_k=1)  &&\text{by (\ref{ass:wA-ignorability_2})} \\
    &= \E( \E(Y | A=0, V_k=1, W, E) \mid V_k=1)  &&\text{by (\ref{ass:cons_2})} \\
    &= \E( \E(Y | A=0, W, E) \mid V_k=1)  &&\text{by (\ref{ass:condex_2})} \\
    &= \E \left( \E \left( \frac{\mathds{1}[A=0] Y}{\P(A=0 | W, E)} | W, E \right) \mid V_k=1 \right)  &&\text{by (\ref{ass:posa0_2})} \\
    &= \frac{1}{\P(V_k = 1)} \E \left( \mathds{1}[V_k=1] \E \left( \frac{\mathds{1}[A=0] Y}{\P(A=0 | W, E)} | W, E \right)\right)  \\
    &= \frac{1}{\P(V_k = 1)} \E \left( \E (\mathds{1}[V_k=1] | W, E ) \E \left( \frac{\mathds{1}[A=0] Y}{\P(A=0 | W, E)} | W, E \right)\right)  &&\text{by (IE)} \\
    &= \frac{1}{\P(V_k = 1)} \E \left( \E \left( \frac{\mathds{1}[A=0] Y \mathds{1}[V_k=1]}{\P(A=0 | W, E)} | W, E \right)\right)  \\
    &= \frac{1}{\P(V_k = 1)} \E \left( \E \left( \frac{\mathds{1}[A=0] Y \E (\mathds{1}[V_k=1] | W, E )}{\P(A=0 | W, E)} | W, E \right)\right) &&\text{by (IE)} \\
    &= \frac{1}{\P(V_k = 1)} \E \left( \E \left( \frac{\mathds{1}[A=0] Y \P (V_k=1 | W, E )}{\P(A=0 | W, E)} | W, E \right)\right) \\
    &= \frac{1}{\P(V_k = 1)} \E \left( \frac{\mathds{1}[A=0] Y \P (V_k=1 | W, E )}{\P(A=0 | W, E)}\right) \\
\end{align*}

\noindent
Note that if $V_k$ is deterministic, then $\P(V_k=1|W,E) = \mathds{1}[E>t]  = \mathds{1}[V_k=1]$ and therefore 

\begin{align*}
   \frac{1}{\P(V_k = 1)} \E \left( \frac{\mathds{1}[A=0] Y \P (V_k=1 | W, E )}{\P(A=0 | W, E)}\right) 
    = \frac{1}{\P(V_k = 1)} \E \left(  \frac{\mathds{1}[A=0]Y\mathds{1}[V_k=1]}{\P(A=0 | W, E)} \right). 
\end{align*}

\end{proof}

\subsection*{Proof of Theorem 2}

\subsection*{Identification of ATE}

Recall that 

\begin{definition}[Conditional and marginal average treatment effect of treatment
  arm $k$ compared to control]
  \begin{align*}
    \cate(k,e,w) &= \E[Y(k) - Y(0)\mid
    W=w,E=e]\\
    \ate(k) &= \E[\cate(k,W,E)].
  \end{align*}
\end{definition}


\subsubsection*{Identification based on the G-formula.}  

\begin{proof}
We show the proof for treatment $0$. 
We refer to $W=w, E=e$ as $W,E$ for clarity.
    \begin{align*}
        \E(Y(0)) &= \E( \E(Y(0) | W, E) ) &&\text{by (IE)} \\
        &= \E( \E(Y(0) | A=0, W, E) ) &&\text{by (\ref{ass:wA-ignorability_2})} \\
        &= \E( \E(Y | A=0, W, E) ) &&\text{by (\ref{ass:cons_2},\ref{ass:posa0_2})} 
    \end{align*}
\noindent
The proof for treatment $k$ can be shown by following the steps for identifying $\E(Y(k) \mid V_k=1)$ in the section above and then assuming (\ref{ass:extra_2}) to be able to marginalize to concurrent and non-concurrent controls combined. Consequently, under 
(\ref{ass:wA-ignorability_2}-\ref{ass:extra_2}), $\cate(k,w,e)$ is identified as \label{stat:3}
   \begin{equation}
    \label{eq:ecateexII}
    \E(Y\mid A=k, W,E, V_k=1) - \E(Y\mid A=0, W, E).    
  \end{equation}
    
\end{proof}

\subsection*{M-estimation details}

We here provide detail on the M-estimation approach for obtaining asymptotic variances for outcome regression and weighted estimators. Recall that $Z_i$ represent the data for the experimental unit $i$, \textit{i.e.}, $Z_i=(E_i, W_i, V_{k,i}, A_i, Y_i)\sim\P$ and consider $X_i=(E_i, W_i)$.


\subsubsection*{Outcome regression}

\paragraph{$\hat{\eate}_{\text{OR}}^{\text{oc}}$} 

This estimator consider only concurrent controls. Let's define $\hat{\eate}_{\text{OR}}^{\text{oc}} = \mu_k - \mu_0$ where $\mu_k$ and $\mu_0$ are the mean outcomes under treatment $k$ and control in the only concurrent control population. We started by considering controls, $\theta_0 = (\beta_0,\mu_0)$ and the following estimating equations
\begin{align*}
    \sum_{i=1}^n h(Z_i,\theta_0) &= \sum_{i=1}^n \begin{pmatrix}
            h_1(Z_i,\beta_0)\\
            h_2(Z_i,\mu_0)
            \end{pmatrix} = 0 
\end{align*}
\noindent 
where $h_1(Z_i,\beta_0) = X_i^\top V_i(1-A_i)(Y_i-X_i\beta_0)$ and $h_2(X_i,\mu_0) = V_i(Z_i\beta_0-\mu_0)$ are the score functions for the model of the conditional mean and the the marginal mean under control, respectively. We consider the following Jacobian matrix of the estimating equations,

\begin{align*}
    \overline{\mathbf{G}}(\hat \theta_0) &= -\frac{1}{n}\sum_{i=1}^n \frac{\partial h(Z_i,\theta_0)}{\partial \theta_0^\top} \biggr\rvert_{\theta_0 = \hat \theta_0} \\
    &= \frac{1}{n}\sum_{i=1}^n \begin{pmatrix}
                \overline{\mathbf{G}}_{11} & \mathbf{0} \\
            \overline{\mathbf{G}}_{21} & \overline{\mathbf{G}}_{22}
            \end{pmatrix} \\
    &= \frac{1}{n}\sum_{i=1}^n \begin{pmatrix}
            X_i^\top V_i(1-A_i)X_i & 0 \\
            -V_iX_i & V_i 
            \end{pmatrix}
\end{align*}
\noindent 
We then constructed the following influence functions 

\begin{align*}
    \varphi(Z_i,\hat \beta_0) &= \overline{\mathbf{G}}_{11}^{-1} h_1(Z_i,\hat \beta_0) \\
    \varphi(Z_i,\hat \mu_0) &= \overline{\mathbf{G}}_{22}^{-1} \left( h_2(Z_i,\mu_0) + (-\overline{\mathbf{G}}_{21}) \varphi(Z_i,\hat \beta_0) \right),
\end{align*}
\noindent
where $\hat \beta_0$ where obtained by ordinary least squares.  We conducted a similar analysis for $\theta_k = (\beta_k, \mu_k)$. Finally, we obtained the variance of $\hat{\eate}_{\text{OR}}^{\text{oc}}$ as, 
\begin{align*}
    \hat V(\hat{\eate}_{\text{OR}}^{\text{oc}}) &= \frac{1}{n} \left( \frac{1}{n} \sum_{i=1}^n \varphi(Z_i,\hat{\eate}_{\text{OR}}^{\text{oc}})\varphi(Z_i,\hat{\eate}_{\text{OR}}^{\text{oc}})^\top  \right),
\end{align*}
\noindent
where $\varphi(Z_i,\hat{\eate}_{\text{OR}}^{\text{oc}})=\varphi(Z_i,\hat \mu_k) - \varphi(Z_i,\hat \mu_0)$.

\paragraph{$\hat{\eate}_{\text{OR}}^{\text{all}}$} 

This estimator considers both concurrent and non concurrent controls when estimating $\E(Y\mid A=0, W=w,E=e)$. Hence, the analysis for $\hat{\eate}_{\text{OR}}^{\text{all}}$ looks the same as that for $\hat{\eate}_{\text{OR}}^{\text{oc}}$ only changing the estimating equation for $\E(Y\mid A=0, W=w,E=e)$, \textit{i.e.}, $h_1(Z_i,\beta_0) = Z_i^\top(1-A_i)(Y_i-Z_i\beta_0)$, while the conditional mean of the outcome among the treated remains computed within only concurrent, \textit{i.e.}, $\E(Y\mid A=k, V_k=1, W=w,E=e)$. Specifically, we started by considering controls, $\theta_0 = (\alpha_0, \mu_0)$ and the following estimating equations
\begin{align*}
    \sum_{i=1}^n h(Z_i,\theta_0) &= \sum_{i=1}^n \begin{pmatrix}
            h_1(Z_i,\alpha_0)\\
            h_2(Z_i,\mu_0)
            \end{pmatrix} = 0 
\end{align*}
\noindent 
where $h_1(Z_i,\alpha_0) = X_i^\top (1-A_i)(Y_i-X_i\alpha_0)$ and $h_2(Z_i,\mu_0) = V_i(X_i\alpha_0-\mu_0)$ are the score functions for the model of the conditional mean and the the marginal mean under control, respectively. While for the treated units we considered, 
$\theta_k = (\beta_k, \mu_k)$ and the following estimating equations
\begin{align*}
    \sum_{i=1}^n h(Z_i,\theta_k) &= \sum_{i=1}^n \begin{pmatrix}
            h_1(Z_i,\beta_k)\\
            h_2(Z_i,\mu_k)
            \end{pmatrix} = 0 
\end{align*}
\noindent 
where $h_1(Z_i,\beta_k) = X_i^\top V_i A_i (Y_i-X_i\beta_k)$ and $h_2(Z_i,\mu_k) = V_i(X_i\beta_k-\mu_k)$. Derivation of the Jacobian matrix of the estimating equations is similar to the above. 

\ignore{

\paragraph{$\hat{\ate}_{\text{OR}}$}. We considered for controls, $\theta_0 = (\alpha_0, \mu_0)$ and the following estimating equations
\begin{align*}
    \sum_{i=1}^n h(Z_i,\theta_0) &= \sum_{i=1}^n \begin{pmatrix}
            h_1(Z_i,\alpha_0)\\
            h_2(Z_i,\mu_0)
            \end{pmatrix} = 0 
\end{align*}
\noindent 
where $h_1(Z_i,\alpha_0) = X_i^\top (1-A_i)(Y_i-X_i\alpha_0)$ and $h_2(Z_i,\mu_0) = (X_i\alpha_0-\mu_0)$ are the score functions for the model of the conditional mean and the the marginal mean under control, respectively. While for the treated units we considered, 
$\theta_k = (\alpha_k, \mu_k)$ and the following estimating equations
\begin{align*}
    \sum_{i=1}^n h(Z_i,\theta_k) &= \sum_{i=1}^n \begin{pmatrix}
            h_1(Z_i,\alpha_k)\\
            h_2(Z_i,\mu_k)
            \end{pmatrix} = 0 
\end{align*}
\noindent 
where $h_1(Z_i,\alpha_k) = Z_i^\top V_i A_i (Y_i-X_i\alpha_k)$ and $h_2(Z_i,\mu_k) = (X_i\alpha_k-\mu_k)$. Derivation of the Jacobian matrix of the estimating equations is similar to the above.
}

\subsubsection*{Parametric inverse probability weighting}

\paragraph{$\hat{\eate}_{\text{IPW}}^{\text{oc}}$}

This estimator consider only concurrent controls. Let's define $\hat{\eate}_{\text{IPW}}^{\text{oc}} = \mu_k - \mu_0$. We started by considering controls, $\theta = (\eta, \mu_0, \mu_1)$ and the following estimating equations
\begin{align*}
    \sum_{i=1}^n h(Z_i,\theta) &= \sum_{i=1}^n \begin{pmatrix}
            h_1(Z_i,\eta)\\
            h_2(Z_i,\mu_0)\\
            h_3(Z_i,\mu_1)
            \end{pmatrix} = 0 
\end{align*}
\noindent 
where $h_1(Z_i,\eta) = X_i^\top V_i(A_i-\pi)$ and $h_2(Z_i,\mu_0) = V_i(\gamma^0_i Y_i-\mu_0)$, $h_3(Z_i,\mu_1) = V_i(\gamma^1_i Y_i-\mu_1)$ are the score functions for the model of the conditional probability and the marginal mean under control and treatment, respectively, and where $\gamma^0_i = \one\{{A_i=0}\}/(1-\pi_i)$, $\gamma^k_i = \one\{{A_i=k}\}/(\pi_i)$, and $\pi_i = \frac{\exp(X_i^\top\eta)}{1+\exp(X_i^\top\eta)}$. We consider the following Jacobian matrix of the estimating equations,

\begin{align*}
    \overline{\mathbf{G}}(\hat \theta_0) &= -\frac{1}{n}\sum_{i=1}^n \frac{\partial h(Z_i,\theta_0)}{\partial \theta_0^\top} \biggr\rvert_{\theta = \hat \theta} \\
    &= \frac{1}{n}\sum_{i=1}^n \begin{pmatrix}
                \overline{\mathbf{G}}_{11} & \mathbf{0} & \mathbf{0} \\
            \overline{\mathbf{G}}_{21} & \overline{\mathbf{G}}_{22} & \mathbf{0} \\
            \overline{\mathbf{G}}_{31} & \mathbf{0} & \overline{\mathbf{G}}_{33} \\
            \end{pmatrix} \\
    &= \frac{1}{n}\sum_{i=1}^n \begin{pmatrix}
            X_i^\top V_i \frac{\exp(X_i^\top\eta)}{1+\exp(X_i^\top\eta)^2} X_i & 0 & 0 \\
            (1-A_i)V_i X_i Y_i \exp(X_i^\top\eta) & V_i & 0 \\
            A_i V_i X_i Y_i \exp(-X_i^\top\eta) & 0 & V_i \\
            \end{pmatrix}
\end{align*}
\noindent 
We then constructed the following influence functions 

\begin{align*}
    \varphi(Z_i,\hat \eta) &= \overline{\mathbf{G}}_{11}^{-1} h_1(Z_i,\hat \eta), \\
    \varphi(Z_i,\hat \mu_0) &= \overline{\mathbf{G}}_{22}^{-1} \left( h_2(Z_i,\mu_0) + (-\overline{\mathbf{G}}_{21}) \varphi(Z_i,\hat \eta) \right), \\
    \varphi(Z_i,\hat \mu_k) &= \overline{\mathbf{G}}_{33}^{-1} \left( h_3(Z_i,\mu_k) + (-\overline{\mathbf{G}}_{31}) \varphi(Z_i,\hat \eta) \right)
\end{align*}
\noindent
where $\hat \eta$ where obtained by ordinary least squares.  Finally, we obtained the variance of $\hat{\eate}_{\text{IPW}}^{\text{oc}}$ as, 
\begin{align*}
    \hat V(\hat{\eate}_{\text{IPW}}^{\text{oc}}) &= \frac{1}{n} \left( \frac{1}{n} \sum_{i=1}^n \varphi(Z_i,\hat{\eate}_{\text{IPW}}^{\text{oc}})\varphi(Z_i,\hat{\eate}_{\text{IPW}}^{\text{oc}})^\top  \right),
\end{align*}
\noindent
where $\varphi(Z_i,\hat{\eate}_{\text{IPW}}^{\text{oc}})=\varphi(Z_i,\hat \mu_k) - \varphi(Z_i,\hat \mu_0)$.

\subsection*{Proof of Theorem 3}

We start by introducing some notation. 
We introduce an operator $\IF: \psi \rightarrow L_2(\mathds{P})$, where $\mathds{P}$ is a probability distribution assumed to lie in some nonparametric model $\mathcal{P}$, that maps functionals $\psi : \mathcal{P} \rightarrow \mathds{R}$ to their influence function $\varphi(z) \in L_2(\mathds{P})$ and where $z$ is our observed data. Recall the following building blocks:

\begin{itemize}
    \item[(bb1)] the influence function of $\mu(x) = \E[Y | X=x]$ is $\IF(\mu(x)) = \frac{\mathds{1}[X=x]}{\P(X=x)}(Y - \E[Y | X=x])$
    \item[(bb2)] the influence function of $\p(x) = \P(X=x)$ is $\IF(p(x)) = (\mathds{1}[X=x] - p(x))$
    \item[(bb3)] $\IF(\psi_1\psi_2) = \IF(\psi_1)\psi_2 + \psi_1 \IF(\psi_2)$ (product rule)
    \item[(bb4)] $\IF(f(\psi)) = \psi' \IF(\psi)$ (chain rule)
    \item[(bb5)] $\P(A,B,C) = \P(A|B,C)\P(B,C)=\P(A|B,C)\P(B|C)\P(C)$
    \item[(bb6)] $\sum_x \mathds{1}[A=k,X=x] = \mathds{1}[A=k]$.
\end{itemize}
\noindent
Finally, recall that the parameter of interest (under the aforementioned identification assumption) is
  \begin{align*}
    \eate(k) &= \E[ \E[Y | A=k, W-w, E=e, V_k = 1] - \E[Y | A=0, W-w, E=e, V_k = 1] \mid V_k=1].
  \end{align*}
 \noindent
 while in the nonparametric model that assumes (\ref{ass:condex}) is 
  \begin{align*}
    \eate(k) &= \E[ \E[Y | A=k, W-w, E=e, V_k = 1] - \E[Y | A=0, W-w, E=e] \mid V_k=1].
  \end{align*}
\noindent
\paragraph{Theorem 3, eq. (4).} We define $(X=x) = (W=w,E=e)$ and pretend that the data is discrete. Recall that under discrete data
\begin{align*}
    \E[\E[Y | A=1, X=x, V_k=1] \mid V_k = 1] &= 
    \frac{\E[ \mathds{1}[V_k=1] \E[Y | A=1, X=x, V_k=1]]}{P(V_k=1)}  \\
    &=\frac{\sum_x\mathds{1}[V_k=1] \E[Y | A=1, X=x, V_k=1] P(X=x)}{P(V_k=1)} \\
    &=\frac{\sum_x\mathds{1}[V_k=1] \mu_{\text{oc}}(1,x,1)p(x)}{P(V_k=1)} \\
    &= \frac{\psi^1_{num}}{\psi_{den}} = \psi^1.
\end{align*}
\noindent
We now analyze the influence function of $\psi^1_{num}$,
\begin{align*}
    &\varphi(Z,\psi^1_{num}) \equiv  \IF\lbrace \psi^1_{num} \rbrace \\ 
    &= \IF \lbrace \sum_x \mathds{1}[V_k=1] \mu_{\text{oc}}(1,x,1) p(x) \rbrace \\
    &= \mathds{1}[V_k=1]\sum_x \left[ \IF \lbrace \mu_{\text{oc}}(1,x,1) \rbrace p(x) + \mu(1,x,1) \IF \lbrace p(x) \rbrace \right] && \text{by (bb3)}  \\
    &= \mathds{1}[V_k=1]\sum_x \left[ \left( \frac{\mathds{1}[A=1,X=x,V_k=1]}{p(1,x,1)}\lbrace Y - \mu_{\text{oc}}(1,x,1) \rbrace \right) \right. p(x) \\ 
    &+ \mu_{\text{oc}}(1,x,1) \left(\mathds{1}[X=x] - p(x) \right) \left. \right] && \text{by (bb1,bb2)} \\
    &= \mathds{1}[V_k=1]\sum_x \left[ \left( \frac{\mathds{1}[A=1,X=x,V_k=1]}{\P(A=1 \mid X=x,V_k=1)\P(V_k=1|X=x)}\lbrace Y - \mu_{\text{oc}}(1,x,1) \rbrace \right) \right. \\ 
    &+ \mu_{\text{oc}}(1,x,1)\mathds{1}[X=x] - \mu_{\text{oc}}(1,x,1) p(x) \left. \right] && \text{by (bb5)} \\
    &= \mathds{1}[V_k=1]\left[ \left( \frac{\mathds{1}[A=1]}{\P(A=1 \mid X=x,V_k=1)}\lbrace Y - \mu_{\text{oc}}(1,x,1) \rbrace \right)  
    + \mu_{\text{oc}}(1,x,1) \right] -  \psi^1_{num}  && \text{by (bb6)} \\
\end{align*}
\noindent
where in the last equality we also used the fact that $\P(V_k=1|X=x) = 1$ under $\mathds{1}[V_k=1]$, and $\psi^1_{num} = \sum_x \mathds{1}[V_k=1]\mu_{\text{oc}}(1,x,1) p(x)$. Analogously we can compute the influence function of $\psi^0_{num}$,
\begin{align*}
    &\varphi(Z,\psi^0_{num}) \equiv  \IF\lbrace \psi^0_{num} \rbrace \\ 
    &= \IF \lbrace \sum_x \mathds{1}[V_k=1] \mu_{\text{oc}}(0,x,1) p(x) \rbrace \\
    &= \mathds{1}[V_k=1]\left[ \left( \frac{\mathds{1}[A=0]}{\P(A=0 \mid X=x,V_k=1)}\lbrace Y - \mu_{\text{oc}}(0,x,1) \rbrace \right)  
    + \mu_{\text{oc}}(0,x,1) \right] -  \psi^0_{num} \\
\end{align*}
\noindent
We now compute the influence function of $\psi_{den}$,
\begin{align*}
    &\varphi(Z,\psi_{den}) \equiv  \IF\lbrace \psi_{den} \rbrace = \mathds{1}[V_k=1] - \psi_{den} \\
\end{align*}
\noindent
We now consider the influence function of $\frac{\psi^1_{num}}{\psi_{den}} = \psi^1$,
\begin{align*}
    &\varphi(Z,\psi^1) \equiv  \IF\lbrace \psi^1 \rbrace  = \frac{\IF \lbrace \psi^1_{num} \rbrace}{\psi_{den}} - \frac{\psi^1_{num}}{\psi_{den}} \frac{\IF \lbrace \psi_{den} \rbrace}{\psi_{den}} \\
    &= \frac{1}{\psi_{den}} \left[  \IF \lbrace \psi^1_{num} \rbrace -  \frac{\psi^1_{num}}{\psi_{den}} \IF \lbrace \psi_{den} \rbrace \right] \\
    &=\frac{1}{\P(V_k=1)} \left[ \left( \mathds{1}[V_k=1]\left[ \left( \frac{\mathds{1}[A=1]}{\P(A=1 \mid X=x,V_k=1)}\lbrace Y - \mu_{\text{oc}}(1,x,1) \rbrace \right)  
    + \mu_{\text{oc}}(1,x,1) \right] -  \psi^1_{num} \right) \right. \\
    &- \left. \frac{\psi^1_{num}}{\psi_{den}} (\mathds{1}[V_k=1] - \psi_{den}) \right]  \\
    &=\frac{1}{\P(V_k=1)} \left[ \left( \mathds{1}[V_k=1]\left[ \left( \frac{\mathds{1}[A=1]}{\P(A=1 \mid X=x,V_k=1)}\lbrace Y - \mu_{\text{oc}}(1,x,1) \rbrace \right)  
    + \mu_{\text{oc}}(1,x,1) \right] -  \psi^1_{num} \right) \right. \\
    &- \left. \frac{\psi^1_{num}}{\psi_{den}}\mathds{1}[V_k=1] + \frac{\psi^1_{num}}{\psi_{den}} \psi_{den} \right]  \\
    &=\frac{\mathds{1}[V_k=1]}{\P(V_k=1)} \left[ \frac{\mathds{1}[A=1]}{\P(A=1 \mid X=x,V_k=1)}\lbrace Y - \mu_{\text{oc}}(1,x,1) \rbrace  + \mu_{\text{oc}}(1,x,1)  -  \psi^1  \right]
\end{align*}
\noindent
We can now combine $\frac{\psi^1_{num} - \psi^0_{num}}{\psi_{den}} = \eate(k)$ to obtain

\begin{align*}
    \varphi(Z,\eate(k)) \equiv  \IF\lbrace \eate(k) \rbrace  
    &=   \frac{\one{\{V_k=1\}}}{\P(V_k=1)}\bigg[\frac{2A-1}{\P(A\mid W, E, V_k=1)}\{Y-\E(Y\mid A, W, E, V_k=1)\}\\
       &+ \E(Y\mid A=1, W, E, V_k=1)-\E(Y\mid A=0, W, E, V_k=1) - \eate(k).\bigg]
\end{align*}

\paragraph{Theorem 3, eq. (5).} Under assumption (\ref{ass:condex}), we now target (among controls),
\begin{align*}
    \E[\E[Y | A=0, X=x] \mid V_k = 1] &= 
    \frac{\E[ \mathds{1}[V_k=1] \E[Y | A=0, X=x]]}{P(V_k=1)}  \\
    &=\frac{\E[ \E[\mathds{1}[V_k=1] \mid X=x] \E[Y | A=0, X=x]]}{P(V_k=1)}  \\
    &=\frac{\sum_x \P(V_k=1 \mid X=x) \E[Y | A=0, X=x] P(X=x)}{P(V_k=1)} \\
    &=\frac{\sum_x \nu(w_i,e_i) \mu_{\text{all}}(1,x)p(x)}{P(V_k=1)} \\
    &= \frac{\psi^0_{num}}{\psi_{den}} = \psi^0.
\end{align*}
\noindent
The influence function of $\psi^0_{num}$ is
\begin{align*}
    &\varphi(Z,\psi^0_{num}) \equiv  \IF\lbrace \psi^0_{num} \rbrace \\
    &=\sum_x \left[ \frac{\mathds{1}[X=x]}{\P(X=x)}(\mathds{1}[V_k=1]-\nu(x)) \mu_{\text{all}}(0,x) p(x) \right. \\ 
    &+  \nu(x) \left( \frac{\mathds{1}[A=0,X=x]}{\P(A=0 \mid X=x)\P(X=x)} \lbrace Y -  \mu_{\text{all}}(0,x) \rbrace \right) p(x)  \\
    &+ \nu(x) \mu_{\text{all}}(0,x) (\mathds{1}[X=x] - p(x)) \left. \frac{ }{ } \right] \\
    &=\sum_x \left[ \frac{\mathds{1}[X=x]\mathds{1}[V_k=1]}{\P(X=x)}\mu_{\text{all}}(0,x)p(x) - \frac{\mathds{1}[X=x]}{\P(X=x)}\nu(x)\mu_{\text{all}}(0,x)p(x)  \right. \\ 
    &+  \nu(x) \left( \frac{\mathds{1}[A=0,X=x]}{\P(A=0 \mid X=x)\P(X=x)} \lbrace Y -  \mu_{\text{all}}(0,x) \rbrace \right) p(x) \\
    &+ \mathds{1}[X=x] \nu(x) \mu_{\text{all}}(0,x)  - \nu(x) \mu_{\text{all}}(0,x) p(x)) \left. \frac{ }{ } \right] \\
    &= \mathds{1}[V_k=1]\mu_{\text{all}}(0,x) + \nu(x) \left( \frac{\mathds{1}[A=0]}{\P(A=0 \mid X=x)} \lbrace Y -  \mu_{\text{all}}(0,x) \rbrace \right) - \psi^0_{num}
\end{align*}
\noindent
As shown before, we can then compute the influence function of $\frac{\psi^0_{num}}{\psi_{den}}$, and finally of $\eate(k)$ under assumption (\ref{ass:condex}), leading to, 

 \begin{align*}
       \varphi(Z,\eate(k)) \equiv \IF \lbrace \eate(k) \rbrace &= \frac{\one{\{V_k=1\}}}{\P(V_k=1)}\bigg[\frac{A}{\P(A\mid V_k=1, W,E)}\{Y-\E(Y\mid A,W,E,V_k=1)\}
       \bigg]\\
       &-\frac{1-A}{\P(A\mid W,E)}\frac{\P(V_k=1\mid E, W)}{\P(V_k=1)}\{Y-\E(Y\mid A, E, W)\}\\
       &+\frac{\one{\{V_k=1\}}}{\P(V_k=1)}\Big[\E(Y\mid A=1,W,E,V_k=1)-\E(Y\mid A=0, W, E)-\eate(k)\Big]
   \end{align*}

\color{black}

\newpage

\subsection*{Additional discussions about platform trials models.}


As one of the anonymous reviewers pointed out, the availability of a treatment may impact the entry time (recruitment rate). If a more ``attractive'' treatment is available, we would expect more participants to be recruited in a shorter period of time, and vice versa. This can be encoded in a DAG and a SCM as an arrow from $V_k$ to $E$, $V_k \rightarrow E$, instead of $V_k \leftarrow E$, and as $E$ as a function of $V_k$, respectively. As shown below, the set $\lbrace V_k,W,E \rbrace$ conditionally blocks the causal path from $A$ to $Y$, leading to the same identification assumptions presented in Section \ref{identification} of the main manuscript. In summary, our proposed identification strategy and estimators remain valid even when entry time is a function of treatment availability.

    \begin{figure}[h!]
        \centering
        \input{dag-b3.tex}
        \caption{\color{black} DAG associated to the structural equation model in equation    (\ref{eq:npsem3}).}     \label{fig:dag-b3}
    \end{figure}
    \begin{figure}[h!]
        \centering
        \color{black}
        \begin{align}
          V_{k,i}&=f_{V_k}(U_V),\notag\\
          E_i&=f_E(V_{k,i},U_{E,i}),\notag\\
          W_i&=f_W(E_i, U_{W,i}),\label{eq:npsem3}\\
          A_i&=f_A(V_{k,i}, W_i, U_{A,i}),\notag\\
          Y_i&=f_{Y}(A_i, W_i, E_i, U_{Y,i}).\notag
        \end{align}
    \end{figure}

\vspace{0.2in}

\noindent 
We study paths from $A$ to $Y$ and then apply d-separation. 
\begin{align*}
    &A \leftarrow V_k \rightarrow E \rightarrow Y && \lbrace V_k \rbrace; \lbrace E \rbrace ; \lbrace V_k,E \rbrace \\
    &A \leftarrow V_k \rightarrow E \rightarrow W \rightarrow Y && \lbrace V_k \rbrace; \lbrace E \rbrace; \lbrace W \rbrace; \lbrace V_k,E \rbrace; \lbrace V_k,W \rbrace; \lbrace E,W \rbrace; \lbrace V_k,W,E \rbrace \\
    &A \leftarrow W  \rightarrow Y && \lbrace W \rbrace \\
    &A \leftarrow W \leftarrow E  \rightarrow Y && \lbrace W \rbrace; \lbrace E \rbrace; \lbrace W,E \rbrace .
\end{align*}
\noindent
By applying d-separation, the set $\lbrace V_k,W,E \rbrace$ conditionally blocks the path from $A$ to $Y$.

\newpage

\subsection*{Additional simulation results.}
\subsection*{Considering a smaller sample size $n=100$.}

\begin{figure}
\begin{center}
\includegraphics[scale=.55]{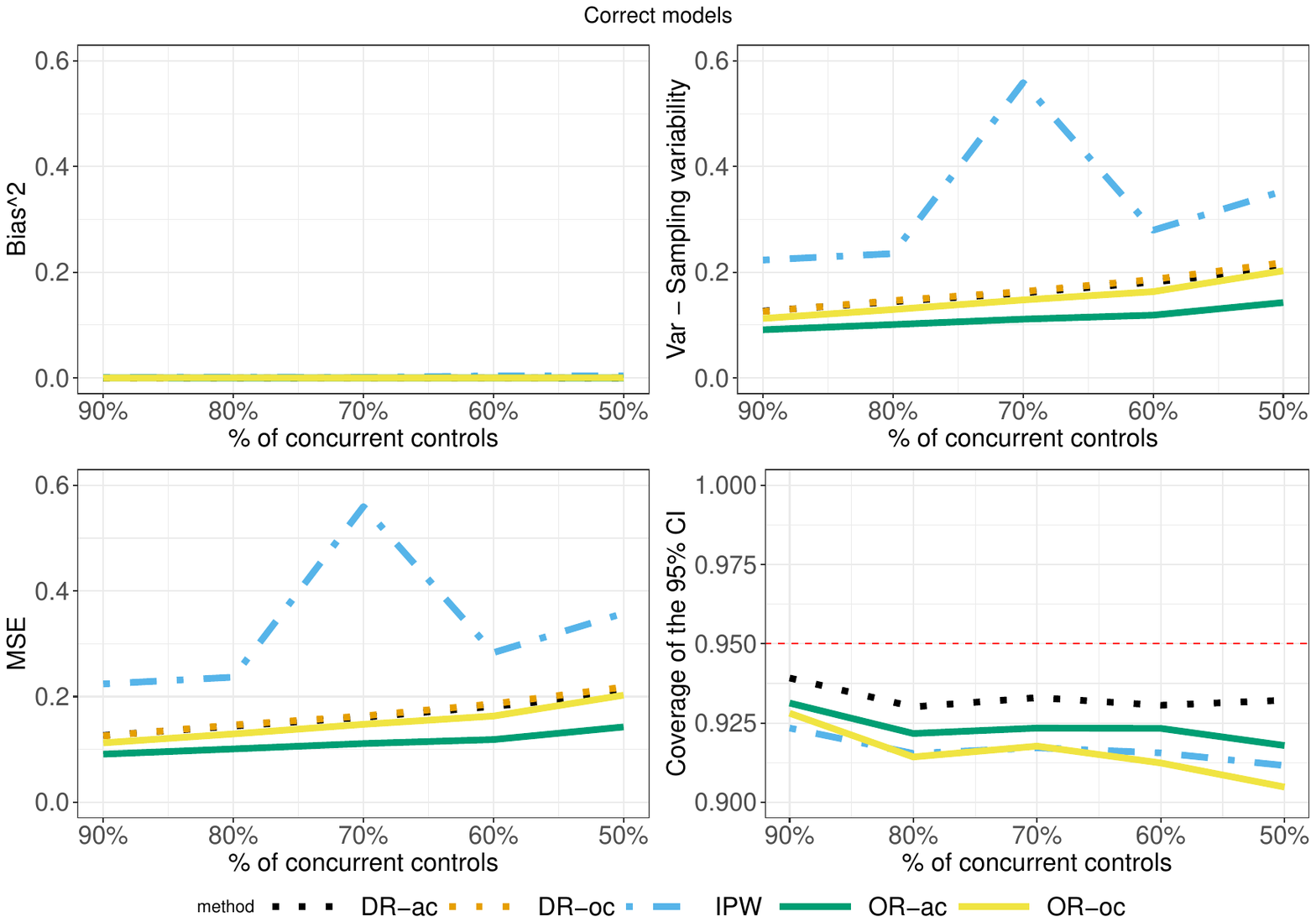}
\end{center}
\caption{ 
\label{fig_corr_corr_n100} \textcolor{black}{$n=100$; Bias squared, variance, MSE and coverage of the 95\% confidence interval of DR-ac, DR-oc, IPW, OR-ac and OR-ac under correct models.  Note that, DR-ac and DR-oc overlap in terms of bias squared, sampling variability and MSE.}}
\end{figure}

\begin{figure}[h]
\begin{center}
\includegraphics[scale=.55]{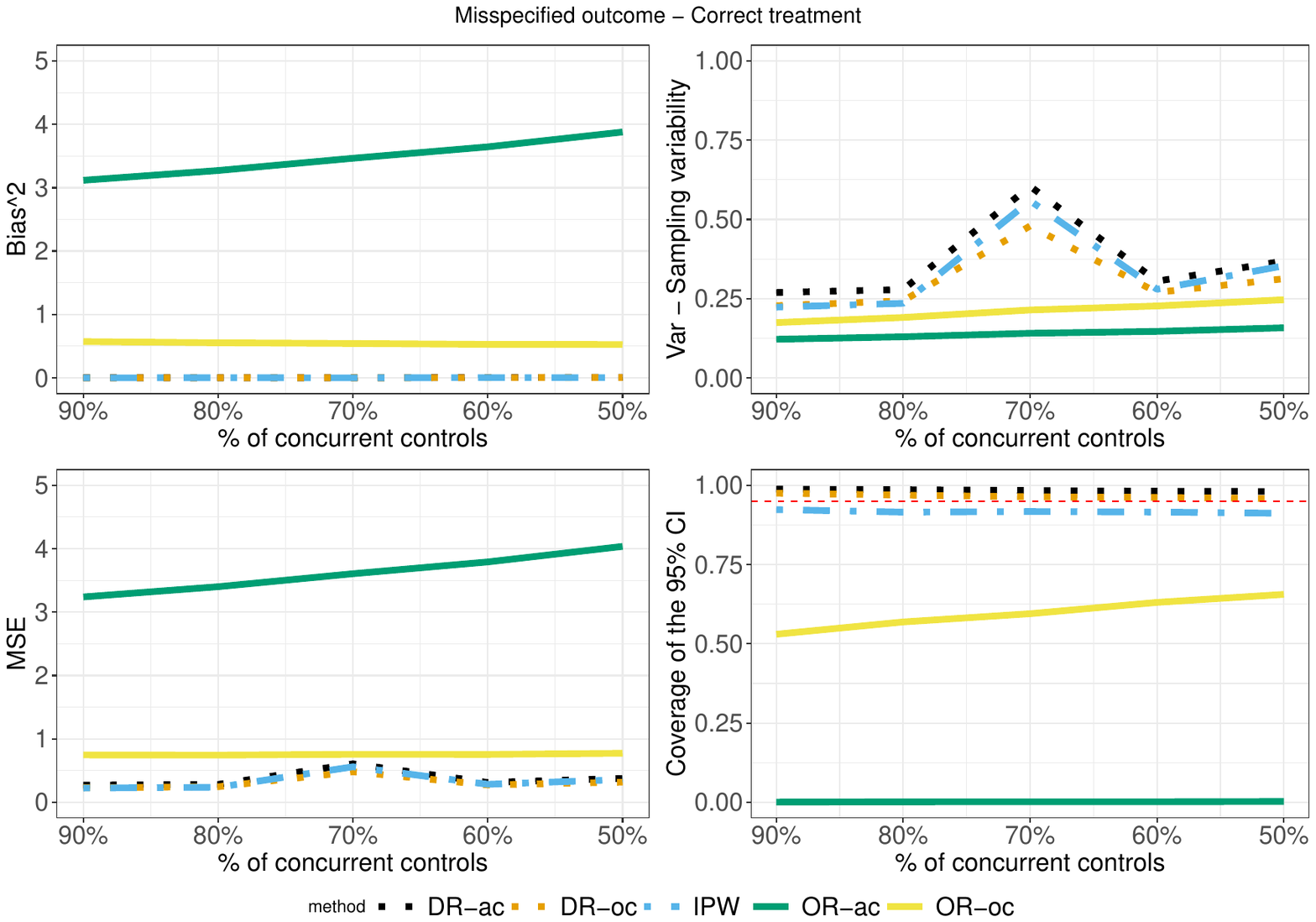}
\end{center}
\caption{ 
\label{fig_miss_corr_n100} \textcolor{black}{$n=100$; Bias squared, variance, MSE and coverage of the 95\% confidence interval of DR-ac, DR-oc, IPW, OR-ac and OR-ac under misspecified outcome model and correct treatment assignment model. Note that, DR-ac and DR-oc overlap in terms of bias squared and MSE.}}
\end{figure}

\begin{figure}
\begin{center}
\includegraphics[scale=.55]{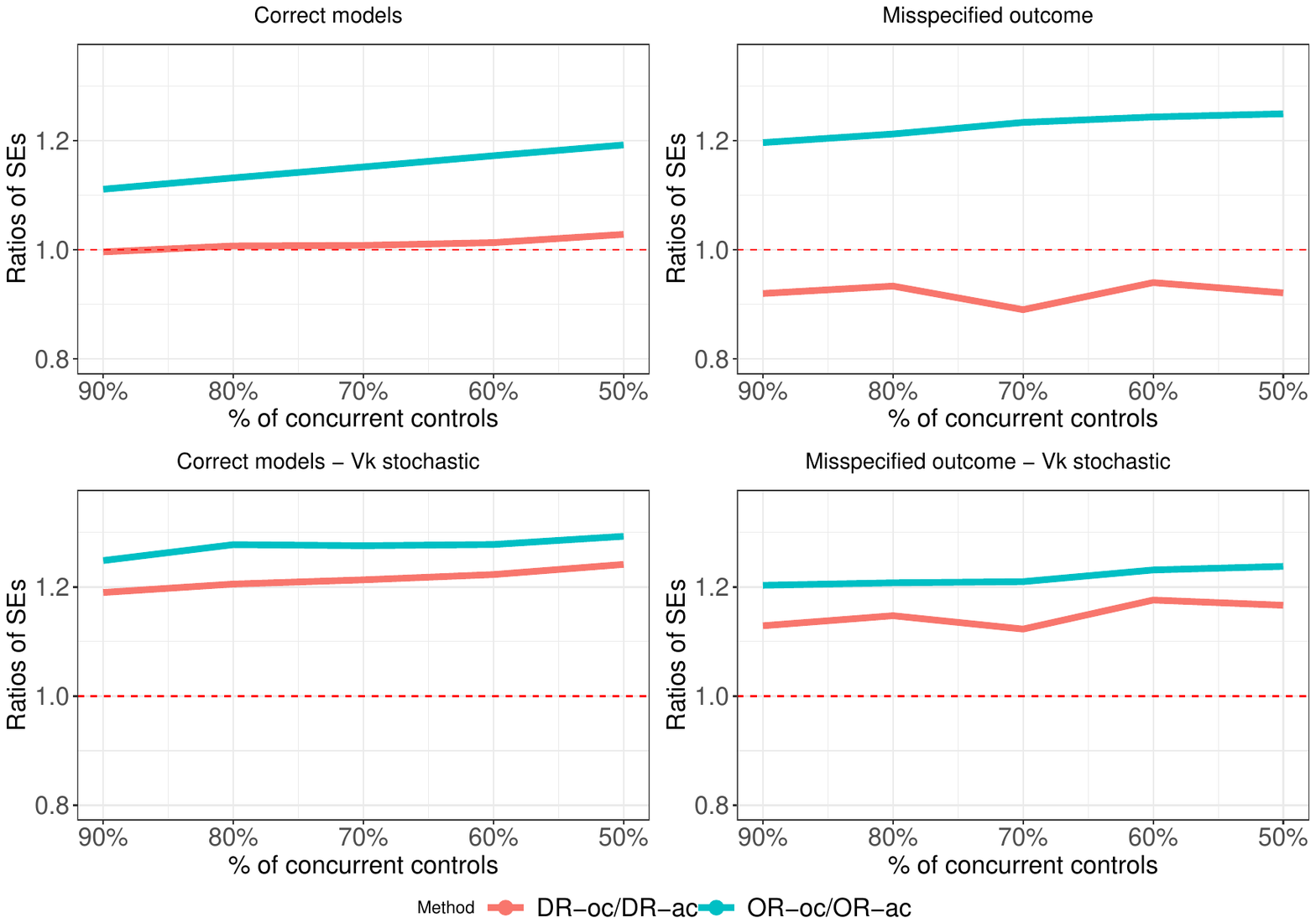}
\end{center}
\caption{ 
\label{fig_ratio_n100} \textcolor{black}{$n=100$; Ratio of standard errors DR-oc/DR-ac and OR-oc/OR-ac across model misspecifications considering $V_k$ a deterministic function of $E$ (top panels) and not (bottom panels). A ratio greater than 1 means efficiency gains.}}
\end{figure}

We here report additional results based on $n=100$. Figures \ref{fig_corr_corr_n100}-\ref{fig_ratio_n100} are based on the same simulation scenarios discussed in section \ref{sec:simu} setting $n=100$. The \% of concurrent controls varied between 90\% to 50\% for numerical reasons given the smaller sample size. 

\newpage

\color{black}
\subsection*{Considering two treatment arms in addition to a shared control arm.}

We provide additional simulation results based on the scenario with two treatment arms and a shared control arm.

\begin{figure}
    \centering
	\includegraphics[scale=0.50]{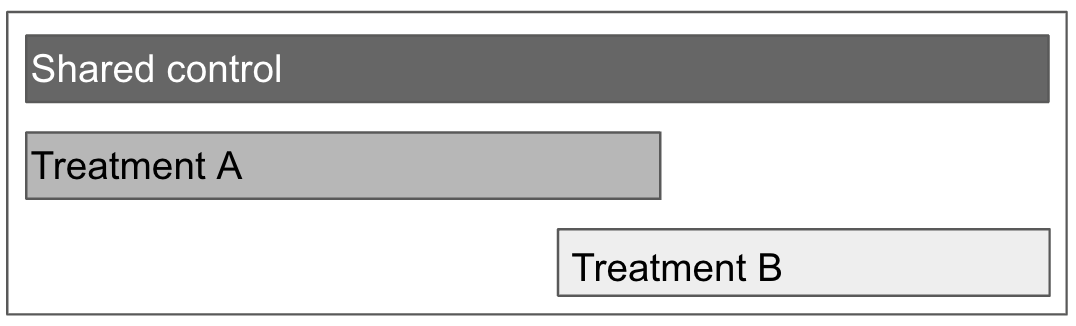}
	\caption{Simulated setting considering one shared control arm and two treatments A and B.}
	\label{fig:trt_k2}
\end{figure}

\paragraph{Aims} 
To evaluate the performance of our proposed estimators when considering two active treatments in addition to a shared control, we specifically considered a scenario where participants were initially randomized to either treatment A or control. Subsequently, treatment B was added, and at some point, treatment A was discontinued. Figure \ref{fig:trt_k2} describes the setting.

\paragraph{Data-generating mechanisms} We considered generating data from Model~(\ref{eq:npsem}). Specifically, we considered a sample size of $n=1,000$ and for each subject $i=1, \dots, n$, we simulated the following data 10,000 times:
\begin{itemize}
    \item[] \textbf{Step 1.} the entry time $E  \sim \text{Norm}(0,1)$ and a baseline covariate $W = -\kappa_1 + 0.8 E + \text{Norm}(0,1)$, where $\kappa_1=n^{-1} \sum_{i=1}^n  0.8 E $;
    \item[] \textbf{Step 2.} two indicators whether treatment $A$ and $B$ were available at time $E$, $V_A$, $V_B$ as a deterministic functions of $E$ being less and greater than a 0.6 and 0.4, respectively. This leads to around 60\% of $V_A=1$, 40\% of $V_B=1$ and around 10\% of $V_A=V_B=1$;
    \item[] \textbf{Step 3.} a 
    \begin{itemize}
        \item[a.] binary treatment $T \sim \text{Bernoulli}(\pi(W))$, where $\pi(W) = \left( 1 + \exp \left( - (-\kappa_2 + 0.8 W) \right)  \right)^{-1}$ and $\kappa_2=n^{-1} \sum_{i=1}^n  0.8 W $ when $V_A=1$ and $V_B=0$, and $A=0$, otherwise  (participants can be randomized to either treatment $A$ or control); 
        \item[b.] binary treatment $T \sim \text{Bernoulli}(\pi(W))$ with $\pi(W)$ as in 3.a when $V_A=0$ and $V_B=1$.
        \item[c.] Multinomial treatment $T \sim \text{Multinom}(\pi_0(W),\pi_A(W),\pi_B(W))$ with $\pi_0(W) = \exp(0)/\tau$, $\pi_A(W) = \pi_B(W) = \exp(\left( - (-\kappa_2 + 0.8 W) \right)))/\tau$ where $\tau = \exp(0) + 2 \exp \left( - (-\kappa_2 + 0.8 W) \right))$ when $V_A=1$ and $V_B=1$.
    \end{itemize}
    
    \item[] \textbf{Step 4.} three counterfactual outcomes, $Y(0) = 0.8 W + 0.5 E + \text{Norm}(0,1)$, $Y(A) = Y(0) + \Delta_A$, and $Y(B) = Y(0) + \Delta_B$ with $\Delta_A=0.5$ and $\Delta_B=1$, and the observed outcome $Y = \one{T=A} Y(A) + \one{T=B} Y(B) + \one{T=0} Y(0)$.
\end{itemize}
\paragraph{Estimands} The estimands of interest are $\E[Y(A) - Y(0) | V_A=1]$, $\E[Y(A) - Y(0) | V_A=1, V_B=0]$, $\E[Y(A) - Y(0) | V_A=1,V_B=1]$, and $\E[Y(B) - Y(0) | V_B=1]$, $\E[Y(B) - Y(0) | V_A=0, V_B=1]$, $\E[Y(B) - Y(0) | V_A=1,V_B=1]$.

\paragraph{Methods} For each dataset, targeted estimand and population, we used the methods summarized in Table \ref{table_methods}. 

\paragraph{Performance metrics} Bias squared, variance, mean square error (MSE), and coverage of the 95\% confidence interval. 

\paragraph{Scenarios} We considered levels of percentage of concurrent controls of 60\% of treatment $A$ 40\% of treatment $B$ and around 10\% for both. We only considered correct models. Specifically, in our simulations, we considered approximately 60\% for $V_A=1$, 40\% for $V_B=1$, 50\% for $V_A=1,V_B=0$, 30\% for $V_A=0,V_B=1$ and 10\% for $V_A=V_B=1$.

\paragraph{Results.}

The tables below provide some results.

\begin{minipage}[t]{0.5\textwidth}
\scriptsize
\begin{tabular}{lrrrrr}
\hline
\textbf{}         & \multicolumn{5}{c}{$\E[Y(A) - Y(0) | V_A=1]$ }                                                                                                                                        \\ \cline{2-6} 
\textbf{}         & \multicolumn{1}{l}{\textbf{DR-ac}} & \multicolumn{1}{l}{\textbf{DR-oc}} & \multicolumn{1}{l}{\textbf{OR-ac}} & \multicolumn{1}{l}{\textbf{OR-oc}} & \multicolumn{1}{l}{\textbf{IPW}} \\
\textbf{bias}     & 0                                  & 0                                  & 0                                  & 0                                  & 0                                \\
\textbf{SE}       & 0.09                               & 0.09                               & 0.08                               & 0.08                               & 0.09                             \\
\textbf{MSE}      & 0.01                               & 0.01                               & 0.01                               & 0.01                               & 0.01                             \\
\textbf{coverage} & 0.95                               & 0.95                               & 0.95                               & 0.95                               & 0.97                             \\
                  & \multicolumn{1}{l}{}               & \multicolumn{1}{l}{}               & \multicolumn{1}{l}{}               & \multicolumn{1}{l}{}               & \multicolumn{1}{l}{}             \\ \cline{2-6} 
                  & \multicolumn{5}{c}{$\E[Y(A) - Y(0) | V_A=1, V_B=0]$}                                                                                                                                 \\ \cline{2-6} 
                  & \multicolumn{1}{l}{\textbf{DR-ac}} & \multicolumn{1}{l}{\textbf{DR-oc}} & \multicolumn{1}{l}{\textbf{OR-ac}} & \multicolumn{1}{l}{\textbf{OR-oc}} & \multicolumn{1}{l}{\textbf{IPW}} \\
 \textbf{bias}                 & 0                                  & 0                                  & 0                                  & 0                                  & 0                                \\
\textbf{SE}     & 0.09                               & 0.09                               & 0.08                               & 0.08                               & 0.1                              \\
\textbf{MSE}       & 0.09                               & 0.09                               & 0.08                               & 0.08                               & 0.11                             \\
\textbf{coverage}      & 0.95                               & 0.95                               & 0.95                               & 0.95                               & 0.96                             \\
 & \multicolumn{1}{l}{}               & \multicolumn{1}{l}{}               & \multicolumn{1}{l}{}               & \multicolumn{1}{l}{}               & \multicolumn{1}{l}{}             \\ \cline{2-6} 
\textbf{}         & \multicolumn{5}{c}{$\E[Y(A) - Y(0) | V_A=1, V_B=1]$}                                                                                                                                 \\ \cline{2-6} 
\textbf{}         & \multicolumn{1}{l}{\textbf{DR-ac}} & \multicolumn{1}{l}{\textbf{DR-oc}} & \multicolumn{1}{l}{\textbf{OR-ac}} & \multicolumn{1}{l}{\textbf{OR-oc}} & \multicolumn{1}{l}{\textbf{IPW}} \\
\textbf{bias}     & 0                                  & 0                                  & 0                                  & 0                                  & 0.01                             \\
\textbf{SE}       & 0.35                               & 0.35                               & 0.24                               & 0.34                               & 0.4                              \\
\textbf{MSE}      & 0.12                               & 0.12                               & 0.06                               & 0.11                               & 0.16                             \\
\textbf{coverage} & 0.94                               & 0.93                               & 0.92                               & 0.92                               & 0.95                             \\ \hline
\end{tabular}
\end{minipage}
\hfill
\begin{minipage}[t]{0.5\textwidth}
\scriptsize
\begin{tabular}{lrrrrr}
\hline
\textbf{}         & \multicolumn{5}{c}{$\E[Y(B) - Y(0) | V_B=1]$}                                                                                                                                        \\ 
\cline{2-6} 
\textbf{}         & \multicolumn{1}{l}{\textbf{DR-ac}} & \multicolumn{1}{l}{\textbf{DR-oc}} & \multicolumn{1}{l}{\textbf{OR-ac}} & \multicolumn{1}{l}{\textbf{OR-oc}} & \multicolumn{1}{l}{\textbf{IPW}} \\
\textbf{bias}     & 0                    & 0                    & 0                    & 0                    & 0.01                 \\
\textbf{SE}       & 0.13                 & 0.13                 & 0.11                 & 0.12                 & 0.14                 \\
\textbf{MSE}      & 0.02                 & 0.02                 & 0.01                 & 0.01                 & 0.02                 \\
\textbf{coverage} & 0.96                 & 0.95                 & 0.95                 & 0.95                 & 0.99                 \\
                  & \multicolumn{1}{l}{}               & \multicolumn{1}{l}{}               & \multicolumn{1}{l}{}               & \multicolumn{1}{l}{}               & \multicolumn{1}{l}{}             \\ \cline{2-6} 
                  & \multicolumn{5}{c}{$\E[Y(B) - Y(0) | V_A=0, V_B=1]$}                                                                                                                                 \\ \cline{2-6} 
                  & \multicolumn{1}{l}{\textbf{DR-ac}} & \multicolumn{1}{l}{\textbf{DR-oc}} & \multicolumn{1}{l}{\textbf{OR-ac}} & \multicolumn{1}{l}{\textbf{OR-oc}} & \multicolumn{1}{l}{\textbf{IPW}} \\
\textbf{bias}     & 0                    & 0                    & 0                    & 0                    & 0                    \\
\textbf{SE}       & 0.14                 & 0.14                 & 0.12                 & 0.13                 & 0.16                 \\
\textbf{MSE}      & 0.02                 & 0.02                 & 0.01                 & 0.02                 & 0.03                 \\
\textbf{coverage} & 0.96                 & 0.96                 & 0.95                 & 0.94                 & 0.99                 \\
                  & \multicolumn{1}{l}{}               & \multicolumn{1}{l}{}               & \multicolumn{1}{l}{}               & \multicolumn{1}{l}{}               & \multicolumn{1}{l}{}             \\ \cline{2-6} 
\textbf{}         & \multicolumn{5}{c}{$\E[Y(B) - Y(0) | V_A=1, V_B=1]$}                                                                                                                                 \\ \cline{2-6} 
\textbf{}         & \multicolumn{1}{l}{\textbf{DR-ac}} & \multicolumn{1}{l}{\textbf{DR-oc}} & \multicolumn{1}{l}{\textbf{OR-ac}} & \multicolumn{1}{l}{\textbf{OR-oc}} & \multicolumn{1}{l}{\textbf{IPW}} \\
\textbf{bias}     & 0                    & 0                    & 0                    & 0                    & 0.01                 \\
\textbf{SE}       & 0.35                 & 0.35                 & 0.24                 & 0.34                 & 0.4                  \\
\textbf{MSE}      & 0.12                 & 0.12                 & 0.06                 & 0.11                 & 0.16                 \\
\textbf{coverage} & 0.95                 & 0.94                 & 0.92                 & 0.92                 & 0.95             \\ \hline
\end{tabular}
\end{minipage}

\end{document}

%% file: dag-b.tex
\begin{tikzpicture}
  \tikzset{line width=1pt, outer sep=0pt,
    ell/.style={draw,fill=white, inner sep=2pt,
      line width=1pt}};
  \node[circle, draw, name=e, ]{$E$};
  \node[circle, draw, name=w, below = 7mm of e]{$W$};
  \node[circle, draw, name=a, below left = 10mm of w]{$A$};
  \node[circle, draw, name=v, above = 5mm of a]{$V_k$};
  \node[circle, draw, name=y, below right = 10mm of w]{$Y$};

  \draw[->](e) to (v);
  \draw[->](e) to (w);
  \draw[->](e) to (y);
  \draw[->](w) to (y);
  \draw[->](w) to (a);
  \draw[->](v) to (a);
  \draw[->](a) to (y);
\end{tikzpicture}

%% file: dag-b2.tex
\begin{tikzpicture}
  \tikzset{line width=1pt, outer sep=0pt,
    ell/.style={draw,fill=white, inner sep=2pt,
      line width=1pt}};
  \node[circle, draw, name=e, ]{$E$};
  \node[circle, draw, name=w, below = 7mm of e]{$W$};
  \node[circle, draw, name=a, below left = 10mm of w]{$A$};
  \node[circle, draw, name=v, above = 5mm of a]{$V_k$};
  \node[circle, draw, name=y, below right = 10mm of w,scale=0.7]{$Y(k)$};

  \draw[->](e) to (v);
  \draw[->](e) to (w);
  \draw[->](e) to (y);
  \draw[->](w) to (y);
  \draw[->](w) to (a);
  \draw[->](v) to (a);
\end{tikzpicture}

%% file: dag-b3.tex
\begin{tikzpicture}
  \tikzset{line width=1pt, outer sep=0pt,
    ell/.style={draw,fill=white, inner sep=2pt,
      line width=1pt}};
  \node[circle, draw, name=v, ]{$V_k$};
  \node[circle, draw, name=e, below = 5mm of v]{$E$};
  \node[circle, draw, name=w, below = 5mm of e]{$W$};
  \node[circle, draw, name=a, below left = 10mm of w]{$A$};
  \node[circle, draw, name=y, below right = 10mm of w]{$Y$};

  \draw[->](v) to (e);
  \draw[->](e) to (w);
  \draw[->, bend left=20](e) to (y);
  \draw[->](w) to (y);
  \draw[->](w) to (a);
  \draw[->, bend right=20] (v) to (a);
  \draw[->](a) to (y);
\end{tikzpicture}